%% file: main.tex
\documentclass[acmsmall,screen]{acmart}
\AtBeginDocument{%
  }

\setcopyright{acmlicensed}
\copyrightyear{2018}
\acmYear{2018}
\acmDOI{XXXXXXX.XXXXXXX}
\acmConference[Conference acronym 'XX]{Make sure to enter the correct
  conference title from your rights confirmation email}{June 03--05,
  2018}{Woodstock, NY}

\acmISBN{978-1-4503-XXXX-X/2018/06}

\usepackage{amsmath}
\usepackage{algorithm}
\usepackage[noend]{algpseudocode}
\usepackage{geometry}
\usepackage{booktabs}
\usepackage{multicol}
\usepackage{float}
\usepackage[dvipsnames]{xcolor}
\usepackage{tcolorbox}
\usepackage{subcaption}
\tcbuselibrary{skins, breakable}
\usepackage{multirow}
\usepackage{makecell}

\definecolor{modifycolor}{RGB}{0, 0, 0}    
\definecolor{rewritecolor}{RGB}{29, 78, 216}   

\definecolor{topatonecolor}{HTML}{dd9f94}      
\definecolor{mrrcolor}{HTML}{f5e8bd}          
\definecolor{avgfivecolor}{HTML}{b0bda0}      

\newcommand{\modify}[1]{\textcolor{modifycolor}{#1}}    

\newtcolorbox{findingbox}[1][]{
  breakable,
  enhanced,
  colback=black!5,
  colframe=black!75,
  boxrule=0.5pt,
  top=3pt,
  bottom=3pt,
  left=6pt,
  right=6pt,
  #1
}

\makeatletter
\newcounter{algocounter}

\makeatother

\begin{document}

\title{Rethinking the Evaluation of Microservice RCA with a Fault Propagation-Aware Benchmark}

\author{Aoyang Fang}
\email{aoyangfang@link.cuhk.edu.cn}
\author{Songhan Zhang}
\email{222010549@link.cuhk.edu.cn}
\author{Yifan Yang}
\email{yifanyang6@link.cuhk.edu.cn}
\author{Haotong Wu}
\email{haotongwu@link.cuhk.edu.cn}
\author{Junjielong Xu}
\email{junjielongxu@link.cuhk.edu.cn}
\author{Xuyang Wang}
\email{xuyangwang@link.cuhk.edu.cn}
\author{Rui Wang}
\email{224040299@link.cuhk.edu.cn}
\author{Manyi Wang}
\email{225045034@link.cuhk.edu.cn}
\author{Qisheng Lu}
\email{qishenglu@link.cuhk.edu.cn}
\author{Pinjia He}
\email{hepinjia@cuhk.edu.cn}
\authornote{Corresponding author.}
\affiliation{%
  \institution{The Chinese University of Hong Kong, Shenzhen}
  \city{Shenzhen}
  \country{China}
}

\renewcommand{\shortauthors}{Fang et al.}

\input{chapters/00abstract.tex}


\keywords{Do, Not, Us, This, Code, Put, the, Correct, Terms, for,
  Your, Paper}

\received{20 February 2007}
\received[revised]{12 March 2009}
\received[accepted]{5 June 2009}

\maketitle

\input{chapters/01introduction}
\input{chapters/02relatedwork}
\input{chapters/03study}

\input{chapters/04method}

\input{chapters/05studydesign}
\input{chapters/06evaluation}

\input{chapters/07discussion}
\input{chapters/09conclusion}

\bibliographystyle{ACM-Reference-Format}
\bibliography{ref}

\end{document}

%% file: chapters/00abstract.tex
\begin{abstract}
While cloud-native microservice architectures have revolutionized software development, their inherent operational complexity makes failure Root Cause Analysis (RCA) a critical yet challenging task.
Numerous data-driven RCA models have been proposed to address this challenge. However, we find that the benchmarks used to evaluate these models are often too simple to reflect real-world scenarios.
Our preliminary study reveals that simple rule-based methods can achieve performance comparable to or even surpassing state-of-the-art (SOTA) models on four widely used public benchmarks.
This finding suggests that the oversimplification of existing benchmarks might lead to an overestimation of the performance of RCA methods.

To further investigate the oversimplification issue, we conduct a systematic analysis of popular public RCA benchmarks, identifying key limitations in their fault injection strategies, call graph structures, and telemetry signal patterns.
Based on these insights, we propose an automated framework for generating more challenging and comprehensive benchmarks that include complex fault propagation scenarios.
Our new dataset contains \modify{1,430} validated failure cases from \modify{9,152} fault injections, covering \modify{25} fault types across \modify{6} categories, dynamic workloads, and hierarchical ground-truth labels that map failures from services down to code-level causes.
Crucially, to ensure the failure cases are relevant to IT operations, each case is validated to have a discernible impact on user-facing SLIs.
    
Our re-evaluation of 11 SOTA models on this new benchmark shows that they achieve low \textit{Top@1} accuracies, averaging 0.21, with the best-performing model reaching merely 0.37, and execution times escalating from seconds to hours.
From this analysis, we identify three critical failure patterns common to current RCA models: \textit{scalability issues}, \textit{observability blind spots}, and \textit{modeling bottlenecks}.
Based on these findings, we provide actionable guidelines for future RCA research.
We emphasize the need for robust algorithms and the co-development of challenging benchmarks.
To facilitate further research, we publicly release our benchmark generation framework, the new dataset, and our implementations of the evaluated SOTA models.
\end{abstract}

%% file: chapters/01introduction.tex
\section{Introduction}

The widespread adoption of cloud-native microservice architectures, while enabling substantial business agility, has also introduced significant operational complexity.
In such distributed systems, failures are not a matter of if, but when~\cite{addeen2019dynamic}.
The complex web of service dependencies means that a single fault can trigger a cascade of failures, propagating rapidly throughout the system.
With enterprise downtime costs estimated to exceed \$23,000 per minute~\cite{cloudzero2025statistics}, the ability to perform rapid and accurate Root Cause Analysis (RCA) has evolved from a technical requirement into a critical business imperative.
Automating the pinpointing of a fault's origin from vast and varied telemetry data (metrics, logs, and traces), which can reduce the Mean Time To Recovery (MTTR), is therefore essential~\cite{zhang2024failure}.
The research community has produced a variety of data-driven RCA approaches.
Early work often concentrated on a single data modality, such as logs~\cite{li2022actionable,pham2024baro,zhang2024illuminating}, metrics~\cite{wang2023interdependent,ikram2022root}, or traces~\cite{gan2023sleuth,ding2023tracediag,tao2024diagnosing,zhang2024trace,jiang2023look,wang2023incremental}.
More recently, a clear trend has emerged towards multi-modal analysis.
These approaches integrate heterogeneous data sources to construct a more holistic system view, promising significantly enhanced diagnostic precision~\cite{sun2024art,lee2023eadro,kong2024enhancing,zhu2024hemirca,sun2025interpretable,wang2024mrca,zheng2024mulan,yu2023nezha,zhang2023robust,gu2023trinityrcl,zhang2024no}.

Despite the apparent sophistication of these models, we find their effectiveness is often evaluated against benchmarks that oversimplify real-world failure scenarios.
This concern is substantiated by our observation that simple rule-based methods can match or even surpass complex state-of-the-art (SOTA) models on four widely-used public benchmarks.
This discrepancy prompts us to ask a fundamental question: Is the reported progress of SOTA algorithms a true reflection of their advanced designs, or is it an artifact of the inherent simplicity of the benchmarks used for their evaluation?
We hypothesize that existing public benchmarks lack the complexity and scale required to meaningfully differentiate the capabilities of sophisticated RCA models.

To rigorously investigate this hypothesis, we conduct a systematic empirical study of popular public benchmarks, examining their fault injection strategies, dependency graph structures, and telemetry signal patterns.
Our analysis reveals a critical limitation: in the majority of cases in existing datasets, the fault symptom of the root-cause service the most prominent.
This characteristic makes the root cause easily identifiable through simple correlation, obviating the need for complex causal inference that advanced models are designed to perform.

Motivated by this clear gap, we argue for the necessity of systematically designed benchmarks that significantly increase the complexity and challenge for RCA algorithms (e.g., introducing more intricate fault propagation patterns).
To this end, this paper proposes a framework that automates the generation, collection, and validation of failure scenarios at scale.
Specifically, our framework incorporates several key innovations over prior work.
(1) \textbf{Systematic fault generation}. Instead of relying on a small set of manually specified injections, our framework systematically explores a vast fault space defined by 31 fault types across 6 categories by layered random sampling. This allows for the generation of diverse and complex failure scenarios, including those that are difficult to orchestrate manually.
(2) \textbf{Impact-driven validation}. Acknowledging that not all injected faults lead to observable issues, we introduce a pragmatic validation oracle that filters for failures causing detectable degradation in user-facing SLIs (e.g., success rate, latency). Through our impact-driven validation process, we identify 25 fault types that consistently produce user-facing degradations, ensuring that our final benchmark contains only operationally relevant failure scenarios. While this approach intentionally focuses on user-impactful failures, which represent a primary concern for SREs, it provides a consistent and automatable standard for dataset quality, a feature absent in prior benchmarks.
(3) \textbf{Highly dynamic system environment}. 
We implement our framework on TrainTicket~\cite{trainticket}, one of the largest open-source microservice systems, which simulates a real-world online ticketing platform with 50 services. To create a more complex and dynamic environment than previous static, low-QPS (Queries Per Second) benchmarks, we subject the system to a dynamic, state-machine-driven workload. Using this approach, we generate a new benchmark dataset comprising 1,430 distinct validated failure cases. This dataset is uniquely annotated with multi-modal telemetry data and a hierarchical root cause labeling scheme. Unlike existing datasets that typically provide one-to-one mappings, our hierarchical labels recognize that a fine-grained failure inherently implies a coarse-grained one. For instance, a function failure necessarily means its containing service is also faulty.

During the dataset construction process, we identify a critical phenomenon often overlooked in existing benchmarks: a large portion of injected faults are silent.
That is, they do not produce any user-facing impact.
This observation highlights the vastness of the potential fault space and underscores the necessity of an impact-driven validation process to create a benchmark focused on operationally relevant problems.
While we initially designed 31 fault types, our validation process revealed that only 25 consistently produced user-facing degradations.
Although this set of 25 fault types may not perfectly mirror the distribution of production incidents, it represents a far broader and more structured exploration of failure modes—including code-level, network, and resource faults—than any existing public dataset.

In the re-evaluation stage, we select 11 recent SOTA RCA methods spanning different data modalities and architectural designs, including single-modal (metrics, traces) and multi-modal approaches.
Rather than simply re-running existing implementations, we systematically re-engineer these methods into a unified evaluation framework with standardized input interfaces.
Our implementation follows modern DevOps practices, containerizing each algorithm and enabling deployment on Kubernetes clusters.
This standardized approach not only ensures fair comparison but also establishes a foundation for the broader RCA research ecosystem, where future datasets adhering to our data format can be seamlessly evaluated across all implemented methods.
On our benchmark, these methods achieve an average \textit{Top@1} accuracy of only 0.21, with the best-performing method reaching merely 0.37, underscoring considerable room for further improvement.
Furthermore, their execution times increased substantially from mere seconds on existing benchmarks to, in some cases, several hours, exposing severe scalability issues.
This dramatic performance degradation reveals deep, systemic weaknesses in current data-driven RCA approaches.
Through systematic failure mode analysis, we identify three critical patterns that expose fundamental limitations in existing models.
(1) \textbf{Scalability issues}. Methods struggle with increased telemetry volume, with execution times escalating from seconds to hours. For example, CausalRCA~\cite{yu2023nezha} requires over 10 minutes to process a single failure case (8 minutes of data) in our benchmark.
(2) \textbf{Observability blind spots}. Models fail when critical services lack sufficient telemetry data, a common scenario in production where monitoring coverage is incomplete.
(3) \textbf{Modeling bottlenecks}. Existing approaches rely on assumptions that break under complex failure conditions, such as stable request rates during incidents.

Based on these findings, we distill actionable guidelines for future RCA research.
A critical insight concerns the gap between academic assumptions of complete observability and the reality of production systems with incomplete telemetry coverage.
We argue that academic benchmarks must proactively simulate these imperfect conditions to foster the development of robust models prepared for production deployment.

To summarize, this paper makes the following contributions:

\begin{itemize}
    \item We are the first to systematically reveal the over-simplicity of widely used public RCA benchmarks, demonstrating that their perceived difficulty is an artifact of simplistic evaluation.
    \item We introduce a novel framework to synthesize a new large-scale, multi-modal benchmark that embodies complex failure phenomena validated by end-user impact.
    \item Through a large-scale re-evaluation, we identify and analyze a set of critical, previously unreported failure patterns in SOTA models, revealing their fragility to complex causality, observability blind spots, and scalability limitations.
    \item We establish a standardized and reproducible evaluation ecosystem for RCA research by re-engineering 11 SOTA algorithms into a unified, containerized framework, facilitating fair and scalable evaluation.
    \item Based on our findings, we distill actionable guidelines for future research and publicly release our entire research artifact, including the benchmark, generation framework, and re-engineered models, to foster community progress.
\end{itemize}

%% file: chapters/02relatedwork.tex
\section{Related Work}

\subsection{Data-Driven Root Cause Analysis in Microservices}

The adoption of microservice architecture has been a paradigm shift in building scalable and resilient cloud applications. 
However, this architectural style introduces significant complexity for system maintenance. 
A single user request may traverse dozens of loosely coupled services, and a fault in one service can propagate unpredictably, causing a cascade of failures across the system. 
This distributed nature makes manual root cause analysis (RCA) a daunting and time-consuming task for engineers~\cite{zhang2024failure}.

To address this challenge, the field has moved towards automated, data-driven RCA. These approaches leverage the vast amount of monitoring data generated by modern cloud systems, which primarily includes metrics, logs, and traces. 
Initial approaches primarily relied on a single data modality, including metric-based methods~\cite{pham2024baro,yu2023cmdiagnostor,li2022actionable,li2022causal,he2022graph,xie2024microservice,lin2024root,ikram2022root}, log-based methods~\cite{pan2021faster,jiang2023look}, and trace-based methods~\cite{yu2021microrank}. 
However, recognizing that a holistic system view is essential, researchers started developing methods that fuse metrics, logs, and traces~\cite{sun2024art,lee2023eadro,sun2025interpretable,yu2023nezha,zhang2024no,zhang2023robust,han2024potential}. 
By correlating information across different data types, these advanced approaches can capture more complex failure patterns, leading to substantial improvements in diagnostic performance.

\subsection{Studies in Root Cause Analysis for Microservices}

A recent comprehensive study by Pham et al.~\cite{pham2024root} provides a critical evaluation of the SOTA in causal inference-based RCA. 
They benchmarked an extensive suite of causal discovery (nine) and RCA methods (twenty-one), concluding that no single algorithm is universally superior, and many struggle to outperform random selection, especially in large-scale systems.
Crucially, their work highlights a fundamental challenge that motivates our own: the performance of methods on existing synthetic datasets does not reliably predict their efficacy on real-world systems.
Pham et al. identified this symptom of inconsistent evaluation results stemming from flawed datasets. 
Our research, therefore, addresses the underlying cause by focusing on the methodological construction of a more faithful and controlled evaluation benchmark. We aim to extend their work by providing a more reliable foundation upon which the true capabilities of RCA algorithms can be more accurately assessed.

%% file: chapters/03study.tex
\section{Preliminary Study}
\label{sec:preliminary_study}


This section presents a preliminary study designed to evaluate the inherent challenges within mainstream public microservices RCA datasets. The primary objective is to scientifically assess the ``simplicity'' of these datasets and demonstrate a causal link between their structural and behavioral characteristics and the ostensibly superior performance of SOTA RCA models. The findings will serve as a foundational justification for the need to construct a new, more realistic, and challenging benchmark. This study is guided by three core research questions (RQs):

\begin{itemize}
    \item \textit{RQ1: What are the primary structural and runtime characteristics of failure scenarios in mainstream public RCA datasets?}
    \item \textit{RQ2: How does a simple heuristic method perform on these datasets compared to SOTA models?}
    \item \textit{RQ3: What explains the strong performance of SimpleRCA on existing benchmarks?}
\end{itemize}

\subsection{Study Design}

\subsubsection{Datasets}

For a comprehensive evaluation, we selected the most commonly used and publicly available multimodal datasets in the microservice RCA domain. These datasets cover a range of systems, including TrainTicket (TT) \cite{trainticket}, OnlineBoutique (OB) \cite{OnlineBoutique}, SockShop (SS) \cite{sockshop}, SocialNetwork (SN) \cite{DeathStarBench}, and MicroSS \cite{gaiadataset}.

\begin{itemize}
\item \textbf{Eadro \cite{eadrodataset}}: Includes two sub-datasets, Eadro-TT and Eadro-SN, designed with a specific focus on network and resource faults.
\item \textbf{Nezha \cite{nezhadataset}}: Includes two sub-datasets, Nezha-TT and Nezha-OB, representing relatively small-scale microservice architectures.
\item \textbf{RCAEval \cite{pham2025rcaeval}}: A benchmark suite Spanning three generations (RE1, RE2, RE3), each including three system scenarios. We didn't include RE1 because it is a unimodal dataset.
\item \textbf{AIOps-2021 \cite{aiopsdataset2}}: The dataset from the AIOps Challenge in 2021, known for its large volume of observational data and various failure types.
\item \textbf{GAIA \cite{gaiadataset}}: A dataset with a large number of fault samples generated from real cloud environment traffic.
\end{itemize}

\subsubsection{Heuristic Method Design (SimpleRCA)}

To address RQ2, we designed \textbf{SimpleRCA}, a heuristic method simulating domain experts' intuitive troubleshooting logic. It takes multi-modal service data as input, uses rule-based anomaly detectors to generate alerts, and identifies the service with the most alerts as the root cause. Tailored rules for each data modality:
\begin{itemize}

\item \textbf{Metrics}: For each service metric, a threshold-based detector is used. The threshold is dynamically determined using robust statistical methods, such as the 3-sigma rule or the 95th percentile (P95) of its historical data. An alert is triggered if a metric's current value exceeds this threshold.

\item \textbf{Traces}: The P95 latency for each service is calculated from its traces. A detector triggers an alert if this latency value is a certain multiple (e.g., 3x) of its normal-period baseline, indicating a significant performance degradation.

\item \textbf{Logs}: A simple keyword-matching approach is employed. The total count of log entries containing common error keywords (e.g., ERROR, FAIL, EXCEPTION) for each service is tallied. An alert is generated if this count crosses a predefined threshold, assuming a sudden spike in errors.

\end{itemize}

These rules are all basic and commonly used anomaly detection rules. SimpleRCA is not a practical RCA solution, but a tool to test whether existing datasets capture real-world complex fault-cause relationships.

\subsubsection{Experimental Setup}
\textbf{SOTA baseline models}: We selected the best-performing methods from the respective benchmark papers (e.g., Eadro, Nezha, Baro) and general-purpose SOTA approaches (e.g., ART for AIOps-2021) for fair comparison.

\textbf{Evaluation metrics}: We utilized a set of widely used evaluation metrics in the RCA domain, including:
\begin{itemize}
\item $Top@K$: The proportion of correct root cause included in the top $K$ predictions.
\item $Avg@K$: The average rank of the correct root cause within the top $K$ predictions.
\item $MRR$ (Mean Reciprocal Rank): The average reciprocal rank for prediction accuracy.
\item Running Time: The time of model execution, evaluating the efficiency of the algorithm.
\end{itemize}

\subsection{Results and Analysis}

\subsubsection{Answering RQ1: Dataset Characteristics}

We first analyzed the static characteristics of existing RCA benchmarks to uncover their inherent structural and runtime properties. Table~\ref{tab:datasets_statistics} summarizes the key statistics, and Fig. ~\ref{fig:rq1_fault_type_distribution} shows the fault type distribution.

\input{tables/rq1_dataset_characteristics.tex}
\input{figures/rq1_fault_type_distribution}

A review of the static features reveals key limitations that simplify the RCA task. \textbf{(1) Limited fault cases.} Except GAIA, all datasets contain fewer than 200 faults (e.g., Eadro and Nezha <100, AIOps-2021 only 159). This scarcity of samples poses a fundamental challenge for complex models to learn robust and generalizable patterns, and the strong performance reported by SOTA methods may largely reflect overfitting to limited scenarios rather than true generalization ability.
\textbf{(2) Simplified request structures.}
Even in datasets with many services (e.g., RE2-TT and RE3-TT with 69 services), all requests collectively cover only 27 services. and has extremely short traces, with seven datasets max out at just 2-3 hops. However, real-world microservice systems typically involve much deeper propagation chains and larger service graphs. These simplified structures reduce the search space and fail to capture the intricate dependencies and cascading failures observed in production environments.
\textbf{(3) Narrow fault spectrum.} Fault types are few and imbalanced. We categorize them into six classes (e.g., CPU, Code, Configuration, Disk, Memory, and Network) based on the fault definitions, but most datasets contain only one or two types (e.g., Eadro sub-datasets and RE3 series) or exhibit severe skew (e.g., GAIA).

\begin{findingbox}{Existing RCA datasets lack realism.}
They feature few fault cases, simplified request structures, and highly imbalanced fault type distributions. These limitations result in benchmarks that hardly approximate real-world complexity, potentially overstating model effectiveness.
\end{findingbox}

\input{tables/rq2_SimpleRCA_Result}
\subsubsection{Answering RQ2: Performance Comparison.}

To address RQ2, we compared SimpleRCA with selected SOTA models on their respective datasets (Table~\ref{tab:simplerca_result}). Baseline results were taken directly from the original papers to ensure faithful and reproducible comparisons.\footnote{We did not include the GAIA dataset in our experiments because it suffers from issues such as inconsistencies in ground-truth annotations (e.g., many injected faults are reflected in the logs, but only a small subset is labeled), which may introduce bias into the evaluation.}

\textbf{Accuracy.} SimpleRCA’s $Top@1$ accuracy is comparable to or surpasses SOTA methods on multiple datasets. For instance, it achieves 0.93 on Nezha-TT versus Nezha’s 0.87, and 0.83 on RE3-TT and RE3-SS versus BARO’s 0.50 and 0.00. 
Even where SOTA models slightly outperform SimpleRCA, the gap is within 0.1, except for Eadro, where the relatively high accuracy is largely due to substantial overlaps in patterns between the training and testing sets.
Furthermore, the advantage of SOTA methods does not widen with relaxed $Top@K$ metrics ($Top@3$, $Top@5$), underscoring the robustness of SimpleRCA across ranking thresholds.

\textbf{Efficiency.} SimpleRCA is orders of magnitude faster than SOTA models, except for the metric-based Baro. However, Baro is a unimodal method that performs far worse than SimpleRCA, with an average Top@1 score on the RCAEVAL dataset that is 0.34 lower (0.24 vs. 0.58). This further emphasizes that on these public benchmark datasets, SOTA models do not offer a performance advantage commensurate with their computational cost.

\begin{findingbox}{SimpleRCA matches or surpasses SOTA performance at far lower cost.}
Despite its simplicity, SimpleRCA attains comparable or superior accuracy to SOTA methods while being orders of magnitude faster, indicating that most faults in existing benchmark datasets can be addressed by simple rule-based methods, making it difficult to effectively evaluate SOTA approaches.
\end{findingbox}

\subsubsection{Answering RQ3: Deficiencies Analysis.}

Based on the fact that SimpleRCA can achieve such superior performance on existing benchmark datasets, we hypothesize that the root cause may lie in deficiencies in the quality of fault injection and observability data within these datasets.
To systematically investigate this issue, we assess the quality of existing datasets along two dimensions: \textbf{(i) completeness of observability data, i.e., whether the key fault-relevant signals are missing, and (ii) simplicity of fault injection patterns, i.e., whether the faults are overly easy to localize}. These two dimensions evaluate data quality in terms of the sufficiency of observational signals and the authenticity and complexity of fault patterns.

Regarding the completeness of observability, we put forward two guiding principles. First, each fault type must be paired with its corresponding observability data (e.g., network faults necessitate full trace coverage; CPU faults require relevant metrics).
Second, from the perspective of user experience, injected faults must propagate to user-facing status codes to reflect real-world disruptions. Otherwise, such faults fail to reflect realistic scenarios, where service disruptions are inevitably observable at the user level. 

\input{tables/rq3_dataset_deficiency}
\input{figures/rq3_case_study}
To assess the simplicity of fault injection patterns, we analyze service-level fault symptoms. A service is deemed to exhibit a \emph{pronounced fault symptoms} if fault-relevant metrics or error counts in logs during the fault period exceed a preset threshold (e.g., twice the average of the preceding normal period); otherwise, it is categorized as showing only an \emph{attenuated fault symptoms}. Based on this rule, each fault case is classified into three types:  
\begin{enumerate}
    \item \textbf{Type I}: Pronounced fault symptoms only in the injected service (i.e., overly localized).
    \item \textbf{Type II}: No pronounced fault symptoms in any service (i.e., fault under-expressed).
    \item  \textbf{Type III}: Stronger fault symptoms in services other than the injected one.  
\end{enumerate}
Type I highlights overly localized fault manifestations, while Type II raises concerns about the effectiveness of fault injection.  
To illustrate, Fig.~\ref{fig:rq3_case_study.tex} shows three CPU faults in Eadro classified as Type I. In these faults, only the injected service exhibits abnormal CPU metrics, while all other services maintain flat trends, making the faults easily identifiable but rarely representative of real-world cascading failures.

Table~\ref{tab:datasets_deficiency} shows that 0.99 of fault cases lack at least one type of observability data, while only four fault cases in the RE2-OB dataset meet our ''comprehensive observability'' criterion.
In terms of fault injection patterns, Type I and Type II together comprise 0.86 of all cases. The prevalence of Type I cases explains why SimpleRCA achieves surprisingly strong performance-faults confined to a single service are inherently easy to localize. By contrast, the existence of Type II cases sheds light on why many algorithms fail, without any pronounced fault signals, localization becomes nearly impossible. Some datasets (e.g., Nezha-TT, Nezha-OB) contain only Type I/II fault cases, suggesting that injected faults are either overly localized or too weak to manifest. Even in RE3-OB, where Type III cases account for 0.7, fault quality remains questionable because most are code-level issues captured in logs, and our classification rule only counts error entries in a straightforward manner.

\begin{findingbox}{Current RCA benchmarks suffer from two systemic shortcomings: (i) incomplete observability depriving models of essential diagnostic signals, and (ii) oversimplified or under-expressed fault patterns making evaluation unreliable. These issues highlight the urgent need for future benchmarks to include diverse, realistic fault scenarios with richer observability signals.}
\end{findingbox}

%% file: tables/rq1_dataset_characteristics.tex
\begin{table*}[t]
\centering
\caption{Key Statistics of Existing RCA Benchmark Datasets. We measure scale by the number of failure cases (\# Cases), services (\# Svcs), total duration, and data volume (logs, traces). Complexity and dynamism are measured by the number of services covered by requests, (\# Svc. by Req), Queries Per Second (QPS), max call depth (Max Depth), metric diversity (\# Metrics) and metric sampling interval (Spl Interval).}
\label{tab:datasets_statistics}
\resizebox{\textwidth}{!}{%
\begin{tabular}{l|rrrrr|rrrrrr}
\toprule
Dataset    & System          & \# Cases & \# Svcs & \# Svc. by Requests & Max Depth & \# Logs (M) & \# Metrics & Spl Interval (s) & \# Traces (M) & Duration & QPS   \\ \midrule
Nezha-TT   & Train-Ticket    & 45       & 46      & 28                  & 4         & 0.16        & 19         & 60               & 0.004         & 8.9hr    & 0.13  \\
Nezha-OB   & Online-Boutique & 56       & 10      & 10                  & 2         & 2.20        & 19         & 60               & 0.047         & 8.1hr    & 1.59  \\
Eadro-TT   & Train-Ticket    & 81       & 27      & 13                  & 3         & 0.01        & 7          & 1                & 0.016         & 17.8hr   & 0.25  \\
Eadro-SN   & SocialNetwork   & 36       & 12      & 12                  & 3         & 0.08        & 7          & 1                & 0.072         & 2.4hr    & 8.38  \\
AIOps-2021 & E-commerce      & 159      & 18      & 12                  & 4         & 5.30        & 483        & 30               & 2.94          & 13.3hr   & 61.63 \\
GAIA       & MicroSS         & 1097     & 10      & 10                  & 2         & 14.60       & 949        & 30               & 3.1           & 31days   & 1.16  \\ \midrule
RE2-TT     & Train-Ticket    & 90       & 69      & 27                  & 5         & 21.30       & 8          & 1                & 0.56          & 36hr     & 4.35  \\
RE3-TT     & Train-Ticket    & 30       & 69      & 27                  & 3         & 1.60        & 8          & 1                & 0.072         & 15hr     & 1.34  \\
RE2-OB     & Online-Boutique & 90       & 17      & 7                   & 2         & 15.00       & 8          & 1                & 2.1M          & 35.7hr   & 16.42 \\
RE3-OB     & Online-Boutique & 30       & 28      & 7                   & 2         & 2.10        & 8          & 1                & 0.28          & 12hr     & 6.51  \\
RE2-SS     & Sock-Shop       & 90       & 16      & NA                  & NA        & 7.60        & 8          & 1                & NA            & 36hr     & NA    \\
RE3-SS     & Sock-Shop       & 30       & 26      & NA                  & NA        & 2.30        & 8          & 1                & NA            & 12hr     & NA    \\ \bottomrule
\end{tabular}
}
\end{table*}

%% file: figures/rq1_fault_type_distribution.tex
\begin{figure}[t]
  \centering
  \includegraphics[width=1\textwidth]{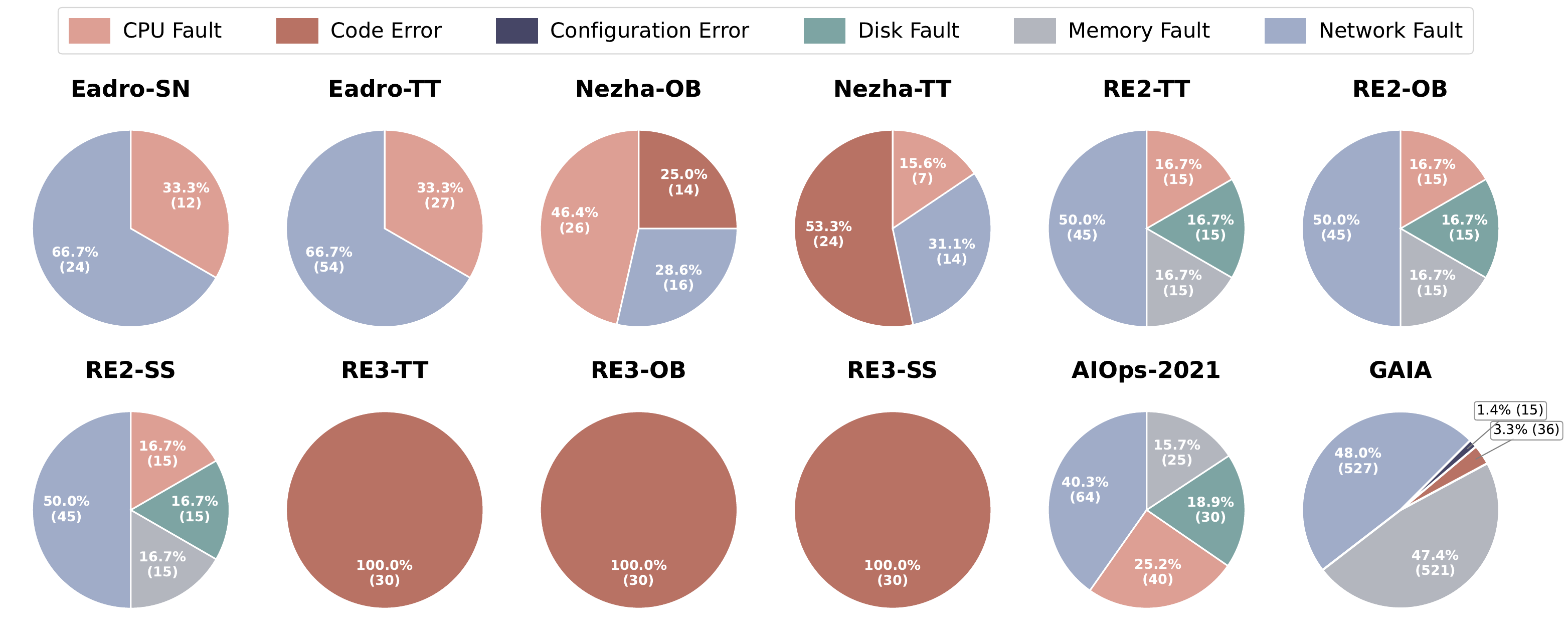}
  \caption{Fault type distribution of existing benchmark datasets.}
  \label{fig:rq1_fault_type_distribution}
\end{figure}

%% file: tables/rq2_SimpleRCA_Result.tex
\begin{table*}[t]
\centering
\caption{Performance Comparison of SimpleRCA and SOTA Methods.}
\label{tab:simplerca_result}
\tiny
\resizebox{0.8\textwidth}{!}{%
\begin{tabular}{ll|rrrrrrr}
\toprule
Dataset    & Method    & $Top@1$ & $Top@3$ & $Top@5$ & $Avg@3$ & $Avg@5$ & MRR  & Time(s) \\ \midrule
\multirow{2}{*}{\centering Re2-OB} & BARO      & 0.14  & 0.93  & 0.98  & 0.64  & 0.77  & 0.54 & \textbf{0.09}    \\
     & \textbf{SimpleRCA} & \textbf{0.47}  & \textbf{0.93}  & \textbf{1.00}  & \textbf{0.68}  & \textbf{0.80}  & 0.67 & 1.14    \\ \midrule
\multirow{2}{*}{\centering RE2-TT}     & BARO      & 0.67  & 0.84  & 0.89  & 0.77  & 0.81  & 0.76 & \textbf{0.87}    \\
     & \textbf{SimpleRCA} & \textbf{0.80}  & \textbf{0.87}  & \textbf{0.93}  & \textbf{0.84}  & \textbf{0.87}  & \textbf{0.85} & 5.99    \\ \midrule
\multirow{2}{*}{\centering RE2-SS}     & BARO      & 0.14  & 0.82  & 0.94  & 0.50  & 0.67  & 0.47 & \textbf{0.06}    \\
     & \textbf{SimpleRCA} & \textbf{0.24}  & \textbf{0.88}  & \textbf{0.97}  & \textbf{0.59}  & \textbf{0.74}  & 0.55 & 0.20    \\ \midrule
\multirow{2}{*}{\centering RE3-OB}     & BARO      & 0.00  & 0.90  & 0.90  & 0.59  & 0.71  & 0.45 & \textbf{0.06}    \\
     & \textbf{SimpleRCA} & \textbf{0.30}  & \textbf{0.90}  & \textbf{0.90}  & \textbf{0.62}  & \textbf{0.73}  & \textbf{0.56} & 0.50    \\ \midrule
\multirow{2}{*}{\centering RE3-TT}     & BARO      & 0.50  & 0.93  & 1.00  & 0.77  & 0.86  & 0.72 & \textbf{0.26}    \\
     & \textbf{SimpleRCA} & \textbf{0.83}  & \textbf{1.00}  & \textbf{1.00}  & \textbf{0.93}  & \textbf{0.96}  & \textbf{0.91} & 1.02    \\ \midrule
\multirow{2}{*}{\centering RE3-SS}     & BARO      & 0.00  & 0.83  & \textbf{0.93}  & 0.46  & 0.65  & 0.40 & \textbf{0.07}    \\
     & \textbf{SimpleRCA} & \textbf{0.83}  & \textbf{0.90}  & 0.90  & \textbf{0.88}  & \textbf{0.89}  & \textbf{0.87} & 0.15    \\ \midrule
\multirow{2}{*}{\centering Nezha-TT}   & Nezha     & 0.87  & 0.98  & 0.98  & NA    & NA    & NA   & NA      \\
   & \textbf{SimpleRCA} & \textbf{0.93}  & \textbf{0.98}  & \textbf{0.98}  & 0.96  & 0.97  & 0.96 & 1.81    \\ \midrule
\multirow{2}{*}{\centering Nezha-OB}   & Nezha     & \textbf{0.93}  & \textbf{0.96}  & \textbf{0.96}  & NA    & NA    & NA   & NA      \\
   & \textbf{SimpleRCA} & 0.91  & 0.91  & 0.91  & 0.91  & 0.91  & 0.91 & 5.05    \\ \midrule
\multirow{2}{*}{\centering Eadro-TT}  & Eadro     & \textbf{0.99}  & \textbf{0.99}  & \textbf{0.99}  & NA    & NA    & NA   & NA      \\
   & \textbf{SimpleRCA} & 0.81  & 0.83  & 0.83  & 0.82  & 0.83  & 0.82 & 0.11    \\ \midrule
\multirow{2}{*}{\centering Eadro-SN}   & Eadro     & \textbf{0.97}  & \textbf{0.99}  & \textbf{0.99}  & NA    & NA    & NA   & NA      \\
   & \textbf{SimpleRCA} & 0.83  & 0.83  & 0.83  & 0.83  & 0.83  & 0.83 & 0.60    \\ \midrule
\multirow{2}{*}{\centering AIOps-2021} & ART       & \textbf{0.72}  & \textbf{0.89}  & NA    & NA    & \textbf{0.87}  & NA   & NA      \\
 & \textbf{SimpleRCA} & 0.63  & 0.82  & 0.91  & 0.74  & 0.80  & 0.74 & 4.25    \\ \bottomrule
\end{tabular}
}
\end{table*}

%% file: tables/rq3_dataset_deficiency.tex
\begin{table*}[t]
\centering
\caption{Statistics on Completeness of Obs ervability and Fault Injection Pattern Classification of Existing Datasets. Note: \# Cases: Count of cases; \# Incom. Cases: Count of cases with incomplete data; R. of Incom. Cases: Rate of incomplete cases; R. of Type $i$: Rate of cases belonging to Type $i$ (i=I,II,III).}
\label{tab:datasets_deficiency}
\scriptsize

\resizebox{\textwidth}{!}{%
\begin{tabular}{l|rrr|rrrr}
\toprule
\multicolumn{1}{c|}{\multirow{2}{*}{\centering \textbf{Dataset}}} & \multicolumn{3}{c|}{\textbf{Completeness of Observability}} & \multicolumn{4}{c}{\textbf{Fault Injection Pattern Classification}} \\  \cmidrule(lr){2-4} \cmidrule(lr){5-8}
\multicolumn{1}{c|}{} & \multicolumn{1}{c|}{\textbf{\# Cases}} & \multicolumn{1}{c|}{\textbf{\# Incom. Cases}} & \multicolumn{1}{c|}{\textbf{R. of Incom. Cases}} & \multicolumn{1}{c|}{\textbf{R. of Type I}} & \multicolumn{1}{c|}{\textbf{R. of Type II}} & \multicolumn{1}{c|}{\textbf{R. of Type III}} & \multicolumn{1}{c}{\textbf{R. of Type I\&II}} \\  
\midrule
Nezha-TT    & 45       & 45              & 1.00               & 0.93         & 0.07         & 0.00         & 1.00             \\
Nezha-OB    & 56       & 56              & 1.00               & 0.91         & 0.09         & 0.00         & 1.00             \\
Eadro-TT    & 81       & 81              & 1.00               & 0.75         & 0.25         & 0.00         & 1.00             \\
Eadro-SN    & 36       & 36              & 1.00               & 0.50         & 0.44         & 0.06         & 0.96             \\
AIOps-2021  & 159      & 159             & 1.00               & 0.63         & 0.16         & 0.21         & 0.79             \\ \midrule
RE2-TT      & 90       & 90              & 1.00               & 0.67         & 0.17         & 0.17         & 0.83             \\
RE3-TT      & 30       & 30              & 1.00               & 0.83         & 0.00         & 0.17         & 0.83             \\
RE2-OB      & 90       & 86              & 0.96               & 0.59         & 0.28         & 0.13         & 0.87             \\
RE3-OB      & 30       & 30              & 1.00               & 0.30         & 0.00         & 0.70         & 0.30             \\
RE2-SS      & 90       & 90              & 1.00               & 0.60         & 0.27         & 0.13         & 0.87             \\
RE3-SS      & 30       & 30              & 1.00               & 0.83         & 0.00         & 0.17         & 0.83             \\
All-Dataset & 737      & 733             & 0.99               & 0.68         & 0.18         & 0.14         & 0.86             \\ \bottomrule
\end{tabular}
}
\end{table*}

%% file: figures/rq3_case_study.tex
\begin{figure}[t]
  \centering
  \includegraphics[width=1\textwidth]{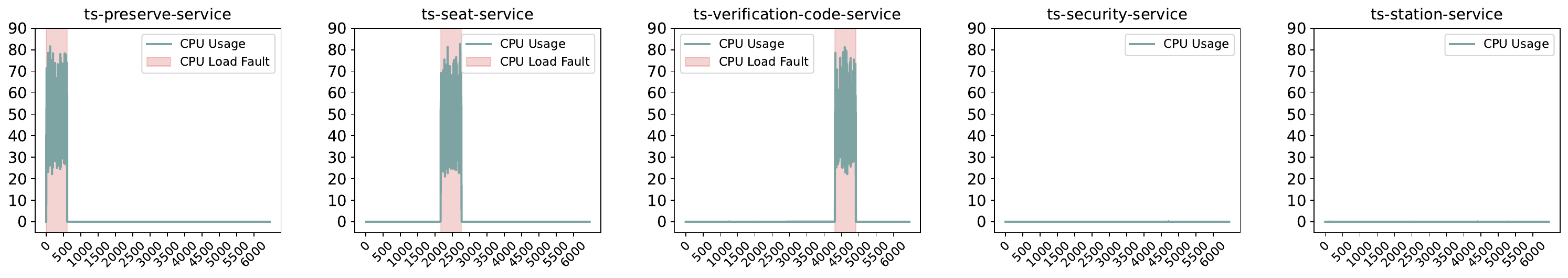}
  \caption{Case Study: CPU load fault in the Eadro dataset. Fault signatures are concentrated only on the injected service; other services, like ts-security-service and ts-station-service, show no observable fluctuations.}
  \label{fig:rq3_case_study.tex}
\end{figure}

%% file: chapters/04method.tex
\section{Dataset Construction}
\label{sec:datasetconstruction}

\subsection{Overview}

In this section, we introduce our systematic framework for constructing a comprehensive benchmark dataset for root cause analysis.
As illustrated in Fig.~\ref{fig:pipeline}, our framework operates as a closed-loop process designed to generate a diverse and impactful set of failure scenarios.
The process begins with a robust \textbf{System Foundation (1)}, the Train-Ticket application, which is subjected to dynamic workloads from our \textbf{Load Generator (2)}.
Concurrently, a \textbf{Controller (3)} orchestrates fault injections into the system using a structured \textbf{Fault Space}, which models a wide array of failure types.
Throughout this process, a multi-modal observability pipeline collects comprehensive telemetry data, including metrics, logs, and traces, via \textbf{OpenTelemetry (4)}.
Crucially, the precise parameters of each injected fault are recorded, forming the basis for our \textbf{Hierarchical Ground Truth Annotation (5)}.
Finally, a \textbf{Validator (6)} automatically assesses the collected data against user-impact metrics (e.g., success rate and latency) to filter out silent faults and categorize the anomalies.
The insights from this validation provide a \textbf{Feedback (7)} mechanism to the controller, enabling a human-in-the-loop strategy to guide subsequent fault injections toward underrepresented and more challenging failure types.
This impact-driven, iterative approach ensures that the resulting dataset is not only large but also diverse and relevant for evaluating modern RCA techniques.

\input{figures/dataset_construction_fig}

\subsection{System Foundation and Enhancements}

The quality of a microservice benchmark is fundamentally constrained by its underlying application. 
Most publicly available systems (e.g., SockShop~\cite{sockshop}, OnlineBoutique~\cite{OnlineBoutique}, SocialNetwork~\cite{DeathStarBench}, MicroSS~\cite{GAIA}) are demonstrators with only about a dozen services, which is difficult to simulate the intricate dependency webs of industrial production environments. 
To overcome this limitation, we selected Train-Ticket~\cite{trainticket}, the largest-to-date (as far as we know) open-source microservice system with \modify{50} distinct services, as our foundation\footnote{Our version is built upon the Nezha~\cite{Nezha} release of Train-Ticket, which instrumented application logs with trace IDs, enabling vital log-trace correlation. }.
We further enhanced this foundation in two key ways:

\begin{itemize}
    \item \textbf{Enhanced Observability}: We deployed a comprehensive monitoring stack on Kubernetes~\cite{kubeprometheus, hubble,otelcollector}, improving the collection of runtime metrics and network telemetry to provide a more holistic view of the system's state.
    \item \textbf{System Stabilization}: We identified and rectified several inherent bugs and performance bottlenecks within the Train-Ticket application.
    For instance, we discovered that the default JVM heap size settings for several services were insufficient.
    Under increased QPS from our dynamic workload generator, these services exhibited non-deterministic latency spikes and instability, contaminating the baseline performance.
    By tuning these configurations and resolving other similar issues, we established a stable system baseline, which is crucial for isolating the impact of deliberately injected faults.
\end{itemize}

\subsection{Dynamic Workload Generation}

\input{figures/loadgenerator}

Previous benchmarks often use static request patterns, which fail to represent the complexity of real-world user interactions.
To address this, we developed a dynamic load generator based on a state-machine model, as illustrated in Fig.~\ref{fig:loadgenerator}.
Our generator models a user workflow as a directed graph where each node represents a \texttt{state} (a required parameter for a request) and each edge represents a \texttt{transition} (a method to fulfill that state).

For instance, a \texttt{Booking Request} is composed of states like \texttt{Passenger Selection} and \texttt{Seat Preference}.
Each state can be resolved through multiple transitions; e.g., \texttt{Passenger Selection} can be done by creating a new user or querying an existing one.
Crucially, transitions can trigger sub-workflows to fulfill prerequisite sub-states, creating a recursive dependency model.
This nested structure leads to a combinatorial explosion of unique execution paths (e.g., the booking workflow has $6 \times 3 \times 2 = 36$ combinations), ensuring a highly dynamic load that mimics real-world user variability.
This approach enables the discovery of defects that static load patterns would miss.
To foster reproducibility, our load generator will be made publicly available.

\subsection{Fault Space Modeling}
\label{sec:fault_space_modeling}

\input{tables/fault_type}

The scope and nature of injected faults are critical for a benchmark's ability to challenge RCA models.
We used ChaosMesh to implement a structured fault space spanning \textbf{31 distinct fault types} across seven categories, as detailed in Table~\ref{tab:fault-types}.
This library includes traditional resource faults as well as more challenging types, such as code-level faults injected via JVM agents and network faults that manipulate HTTP semantics.

We discretize continuous parameters (e.g., CPU cores, latency intervals) to transform the abstract fault space into a concrete, enumerable set of experiments.
The resulting fault space exceeds $10^{16}$ possible configurations, with fine-grained discretization to capture subtle parameter-dependent faults.

\subsection{Multi-modal Data Collection}
We employed a standard observability stack based on OpenTelemetry to collect multi-modal data, including Kubernetes-level metrics, trace-correlated logs, distributed traces, and L4/L7 network telemetry.
All data is normalized into a unified schema and stored in ClickHouse for efficient querying.
This standardized schema, publicly available in our repository, simplifies data consumption for downstream RCA algorithms.

\subsection{Hierarchical Ground Truth Annotation}

A precise ground truth is essential for evaluating RCA models.
In our framework, the ground truth for each failure case is directly derived from the fault injection parameters.
This includes the specific fault type, its target (e.g., service, pod), and ensuring unambiguous root cause labels even in cases of complex, cascading failures.
Previous studies have used different granularities for root cause labels, ranging from coarse service-level identification to fine-grained function-level pinpointing~\cite{yu2023nezha,wang2024mrca}, however, we argue that a more fine-grained label can contain the information of coarse-grained labels, while the reverse is not true. For example, knowing that the root cause is a specific function failure inherently implies that the parent service is also faulty, but not vice versa.

As a result, to support evaluation at varying levels of diagnostic precision, we introduce a hierarchical labeling scheme.
This hierarchy is represented by a tree structure, as shown in Fig.~\ref{fig:label}.
Each failure case is annotated with a set of labels corresponding to these levels of granularity: Service, Pod, Container, and where applicable, more specific identifiers for Metrics, Spans, or Functions.
This structure provides researchers with the flexibility to evaluate their models at different levels of precision, from coarse service-level localization to fine-grained function-level identification\footnote{In this study, we only use service level labels, since it is the only common granularity across all methods.}.

\subsection{Impact-Driven Selection}
\label{sec:impact_selection}

As discussed in preliminary study~\ref{sec:preliminary_study}, a significant methodological flaw in existing benchmarks is the often-unspoken assumption that every injected fault manifests as an observable problem.
In practice, many faults are silent or their effects are absorbed by system resilience mechanisms.
To construct a benchmark of operationally relevant failures, we implement a strict, impact-driven selection process where a fault is included if and only if it produces a discernible anomaly at the system's entry points, as measured by user-facing Service Level Indicators (SLIs).

Our automated validation logic employs a hybrid approach to identify anomalies.
For success rate degradations, it uses Z-tests to detect statistically significant drops.
For latency anomalies, it combines hard thresholds (e.g., 10 seconds) with adaptive statistical thresholds based on historical performance patterns.
This process classifies experiments into two categories: \textbf{Has Anomaly} (experiments with validated user-perceivable impact) and \textbf{No Anomaly} (filtered out from the final dataset).
We acknowledge this is a trade-off that prioritizes operational relevance and scalable validation over the inclusion of subtle "gray failures," which we discuss further in Section~\ref{sec:implications}.

\subsection{Fault Injection Strategy}
\label{sec:injection_strategy}

Given the vastness of the fault space (exceeding $10^{16}$ configurations), an exhaustive exploration is computationally infeasible.
For the construction of the current dataset, we adopted a stratified random sampling strategy.
We first partitioned the entire fault space into strata based on the seven high-level fault categories defined in Table~\ref{tab:fault-types}.
Then, we performed random sampling of fault parameters within each stratum.
This approach ensures that all major fault categories are represented in the dataset, providing a broad foundation for evaluation.

%% file: figures/dataset_construction_fig.tex
\begin{figure}[t]
\centering
  \includegraphics[width=0.8\textwidth]{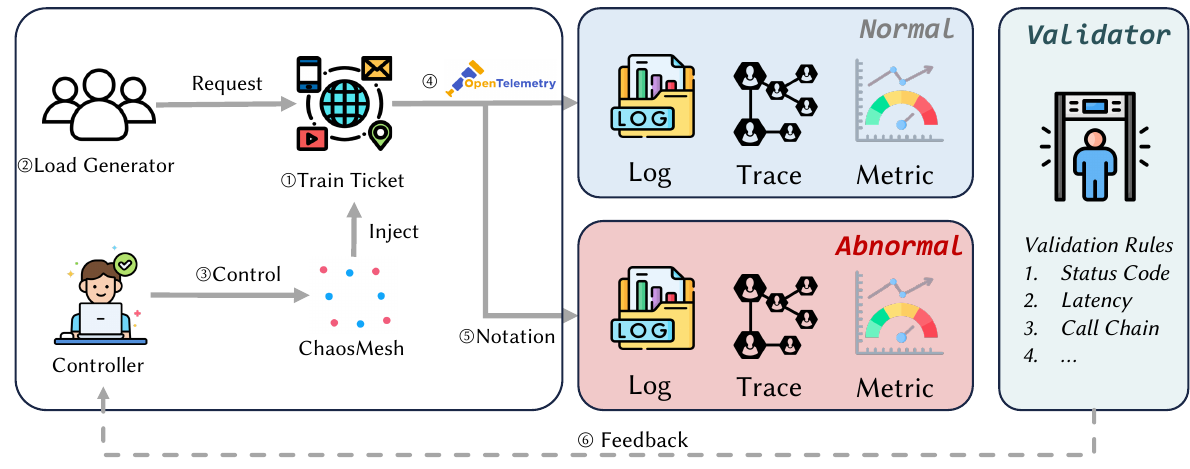}
  \caption{The six-stage pipeline for our benchmark dataset construction, from system selection to the final validated and labeled failure cases.}
  \label{fig:pipeline}
\end{figure}

%% file: figures/loadgenerator.tex
\begin{figure}[t]
  \centering
  \includegraphics[width=0.8\textwidth]{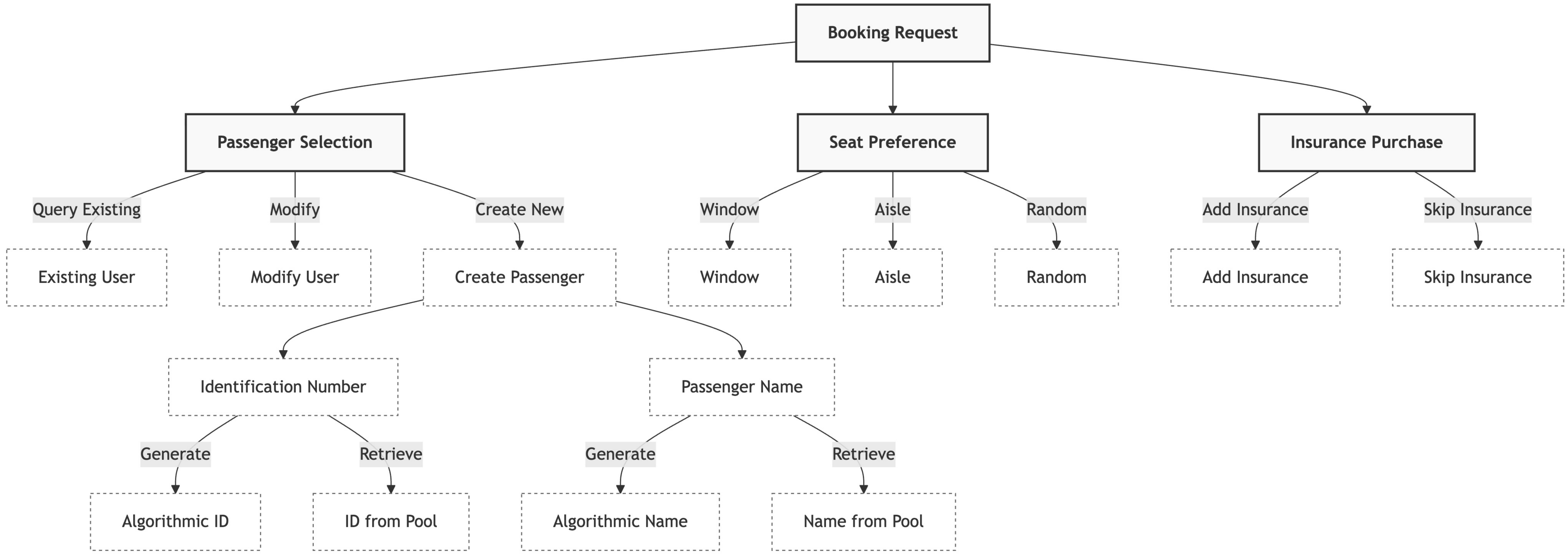}
  \caption{State-machine-based workflow generation model, the booking request node depends three parameters, i.e., passenger, seat, and insurance. And each of the parameter can be fetched in different ways (either from creating new one, or query existing one). This recursive way leads to the combinational explosion.}
  \label{fig:loadgenerator}
\end{figure}

%% file: tables/fault_type.tex
\begin{figure}[t]
\centering
\begin{minipage}{0.6\textwidth}
\centering
\captionsetup{type=table}
\caption{The 31 Fault Types Supported in Our Benchmark, Organized by Category.}
\small
\label{tab:fault-types}
\begin{tabular}{@{}ll@{}}
\toprule
\textbf{Category} & \textbf{Fault Types} \\ \midrule
\textbf{Resource} & PodKill, PodFailure, ContainerKill, MemoryStress, CPUStress \\
\textbf{Network} & Delay, Loss, Duplicate, Corrupt, Bandwidth, Partition \\
\textbf{HTTP} & ReqAbort, RespAbort, ReqDelay, RespDelay, \\
& RespReplaceBody, RespPatchBody, ReqReplacePath, \\
& ReqReplaceMethod, RespReplaceCode \\
\textbf{Code (JVM)} & Latency, Return, Exception, GC, CPUStress, \\
& MemoryStress, MySQLLatency, MySQLException \\
\textbf{DNS} & DNSError, DNSRandom \\
\textbf{Time} & TimeSkew \\ \bottomrule
\end{tabular}
\end{minipage}
\hfill
\begin{minipage}{0.36\textwidth}
\centering
\includegraphics[height=3.5cm]{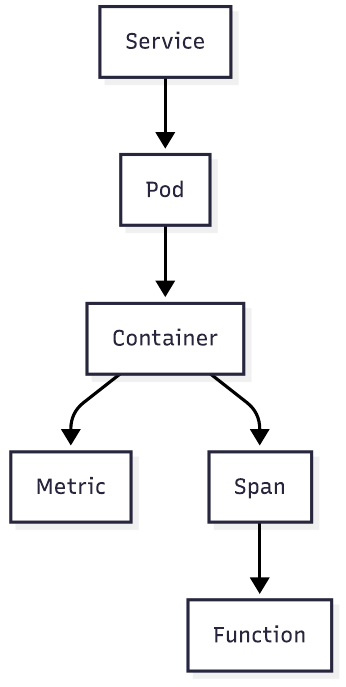}
\caption{The hierarchical structure of our ground truth labels. A fault can be localized at different levels of granularity, from the service down to the specific function.}
\label{fig:label}
\end{minipage}
\end{figure}

%% file: chapters/05studydesign.tex
\section{Study Design}
\label{sec:studydesign}

Based on our preliminary study (Section~\ref{sec:preliminary_study}), which revealed the inherent simplicity of existing RCA benchmarks, this section details the design of our main empirical study.
The primary goal is to introduce and validate a more challenging benchmark and re-evaluate the true capabilities of SOTA models against it.
This study is guided by two core research questions (RQs), which address the quantitative complexity of our new benchmark and the performance of existing models on it.

\subsection{Research Questions}

\paragraph{RQ4} \textit{How does our newly constructed benchmark quantitatively differ from existing public benchmarks in terms of scale, diversity, and complexity?}
To answer this, we will first present statistics from our dataset construction process to demonstrate the efficacy of our methodology.
We will then introduce a set of indicators to quantify and compare key characteristics across benchmarks, including dependency graph complexity, workload dynamics, and the distribution of failure signatures.
This analysis will substantiate our claim that the new benchmark represents a more diverse and challenging set of failure scenarios.

\paragraph{RQ5} \textit{How do SOTA RCA models perform on this new benchmark, and what are their primary failure modes when confronted with complex causality and incomplete observability?}
Through this question, we intend to probe the limitations of current SOTA models using our validated, challenging benchmark.
We hypothesize that the increased complexity will lead to a significant performance degradation.
By analyzing the failure modes against specific, engineered challenges, we aim to provide clear, actionable directions for future research in the field.

\subsection{Evaluation Metrics}
\label{sec:metrics}

To evaluate model performance on our new benchmark, we adopt the same evaluation metrics established in our preliminary study (Section~\ref{sec:preliminary_study}), which include effectiveness metrics (\textit{Top@K}, \textit{Avg@K}, \textit{MRR}) and efficiency metrics (\textit{Execution Time}).
These metrics together provide a comprehensive and standardized view of model performance, enabling direct comparison between our findings on simple and complex benchmarks.

Additionally, to assess the scalability characteristics of different approaches, we conduct a dedicated scalability experiment that evaluates execution time across varying trace volumes, ranging from 2,000 to 18,000 traces.
This experiment specifically focuses on models that utilize trace and log data, as these data types exhibit dramatic volume fluctuations in response to request patterns, unlike metrics data, which typically scales linearly with time.
The scalability analysis provides crucial insights into the practical deployment considerations of each model under realistic workload variations.

\subsection{Evaluation Setup}
\label{sec:setup}

Our evaluation setup is designed to provide a rigorous and reproducible assessment of RCA models.

\paragraph{Baselines}
To ensure a thorough and representative comparison, we evaluate against eleven recent SOTA RCA models that have been published in top-tier software engineering and systems venues.
These models were selected to cover a wide spectrum of underlying techniques in the field.
Specifically, our selection includes:
\begin{itemize}
    \item (1) Graph-based models, such as \textit{MicroRank}~\cite{yu2021microrank}, \textit{MicroRCA}~\cite{wu2020microrca}, \textit{MicroHECL}~\cite{liu2021microhecl}, and \textit{MicroDig}~\cite{tao2024diagnosing}, which leverage service dependency graphs and random walk algorithms to pinpoint faulty services.
    \item (2) Machine learning and deep learning approaches, such as \textit{DiagFusion}~\cite{zhang2023robust}, \textit{Eadro}~\cite{lee2023eadro}, and \textit{Art}~\cite{sun2024art}, which learn complex failure patterns from historical and multimodal data using techniques like GNNs and Transformers.
    \item (3) Causal inference and statistical methods, including \textit{Shapleyiq}~\cite{li2023shapleyiq}, \textit{Nezha}~\cite{yu2023nezha}, \textit{CausalRCA}~\cite{xin2023causalrca}, and \textit{Baro}~\cite{pham2024baro}, which aim to identify the causal contributors to a failure through techniques like Shapley values, event pattern mining, and hypothesis testing.
\end{itemize}

By comparing against this diverse set of baselines, we can draw more robust conclusions about the true state of the art and its limitations when facing complex failure scenarios.
To ensure a fair comparison, we conducted a systematic, best-effort hyperparameter tuning for each model to adapt it to the unique characteristics of our new, more complex benchmark, rather than simply adopting the default configurations provided in the original papers, which were optimized for simpler, smaller-scale datasets.

\paragraph{Dataset}
We conduct our evaluation on the new benchmark dataset proposed in this paper (detailed in Section~\ref{sec:datasetconstruction}).
This benchmark is designed to challenge SOTA models through its dynamic workloads, diverse fault types, and complex causal propagation chains.
Our dataset construction process follows a rigorous protocol: each data collection cycle consists of 4 minutes of normal operation data followed by 4 minutes of failure data.
After each fault injection, the entire service environment is redeployed to ensure complete isolation and prevent interference from previous experiments.
Before each normal data collection phase begins, the system undergoes a 4-minute warm-up period to reach a stable operational state, ensuring the reliability and consistency of our baseline measurements.
For models that require training, such as Art~\cite{sun2024art}, DiagFusion~\cite{zhang2023robust}, and Eadro~\cite{lee2023eadro}, we partition the dataset by fault type, allocating 80\% for training and the remaining 20\% for testing.

\paragraph{Environments}
All experiments are conducted on a server equipped with two EPYC 9754 CPUs and 1.48TiB of DDR5 memory.
For the dataset construction phase, we deploy a Kubernetes cluster consisting of 1 master node and 6 worker nodes, where each worker node is equipped with 128 CPU cores and 128GB of memory.
For algorithm evaluation, each algorithm is executed within a controlled environment, constrained to 128 CPU cores and 128GB of memory.
This setup guarantees that performance differences are attributable to the models' intrinsic capabilities rather than variations in computational resources.

%% file: chapters/06evaluation.tex
\section{Study Results}

This section presents the results of our main empirical study, guided by the research questions established in Section~\ref{sec:studydesign}.
We first answer \textit{RQ4} by quantitatively demonstrating our benchmark's increased scale and complexity compared to prior work.
We then address \textit{RQ5} by re-evaluating eleven SOTA models on this new benchmark, analyzing their performance degradation and systematically identifying three primary failure modes that reveal systemic weaknesses in current RCA approaches.
This analysis provides a foundation for the actionable insights discussed in Section~\ref{sec:discussion}.

\subsection{RQ4: New Dataset Characteristics}

To quantitatively demonstrate the increased complexity and scale of our new benchmark, we first show the effectiveness of our impact-driven data generation process and then conduct a multi-faceted comparative analysis against prior work.
Our evaluation centers on three critical dimensions of a benchmark's complexity and diagnostic challenge: \textbf{Workload Intensity}, \textbf{System Interaction Coverage}, and \textbf{Causal Complexity}.

\subsubsection{Effectiveness of the Impact-Driven Generation Process}
\label{sec:rq4_generation_effectiveness}

\input{figures/rq4_generation_process.tex}

To construct a benchmark with genuine diagnostic challenges, we filtered out inconsequential fault injections through a rigorous, impact-driven selection process described in Section~\ref{sec:impact_selection}.
Fig.~\ref{fig:rq4_generation_process} presents the outcome of this process across over \modify{9,152} initial fault injection experiments spanning \modify{31} distinct fault types, from which we ultimately derived our \modify{1,430} validated failure cases.
The distribution of these experiments across fault types is intentionally non-uniform, reflecting the intrinsic differences in the "injection space" for each fault category.
For instance, JVM-related faults encompass a vast number of code methods as potential injection targets, whereas \texttt{PodKill} faults are limited to the much smaller number of available pods.
To preserve the uniqueness and integrity of each experiment, we did not artificially balance these counts through duplicate injections; each experiment represents a distinct fault configuration.

\begin{findingbox}{Effective fault injection requires careful calibration of both injection intensity and injection location to avoid undetectable weak faults, unreachable code paths, and trivially obvious strong faults.}
\end{findingbox}

Our analysis reveals that a vast majority of initial fault injections \modify{(84.4\%) } were classified as ``No Anomaly'', failing to produce any user-perceivable impact.
This high failure rate is not random but stems from two orthogonal and equally critical dimensions: injection intensity and injection location.

First, \textbf{resource-based fault injection requires delicate calibration within a narrow effective range}.
ChaosMesh implements resource stress faults by spawning independent processes that compete for system resources.
For \texttt{CPUStress} and \texttt{MemoryStress} faults, near \modify{100\% of} our initial injections produced no impact due to intensity miscalibration.
When stress levels are too low, the target application continues operating normally, rendering the injection ineffective.
Conversely, when stress levels are too high, the system's out-of-memory killer or scheduler intervenes to terminate the stress process, again nullifying the injection.
This narrow effective window makes resource-based faults particularly challenging to calibrate and explains their high failure rate in producing meaningful diagnostic scenarios.

Second, \textbf{when injection location is poorly chosen, even high-intensity faults may never manifest}.
This issue was most prominent in JVM-level manipulations, such as \texttt{JVMReturn} and \texttt{JVMCPUStress}.
The primary reason for their failure was that the targeted methods were never invoked by our workload generator, rendering the fault injection inert regardless of its configured intensity.
This dual failure mode validates our premise that many existing benchmarks, which often use minimal intensities and naive location selection, fail to generate meaningful diagnostic challenges.

In contrast, our process also identified faults that consistently produced clear and severe failures.
Manipulations of core HTTP semantics, such as \texttt{HTTPResponseReplaceCode}\modify{ (98\%)} and \texttt{HTTPRequestAbort} \modify{(100\%)}, predictably led to ``Has Anomaly'' classifications.
This is because these faults directly disrupt the fundamental communication protocols that microservices rely on, leading to immediate and obvious user-facing errors.
After filtering out the inconsequential injections, \modify{6} fault types are filtered out, since they never produced any impactful failures in our experiments.
The final benchmark thus consists of \modify{1,430} unique fault configurations across \modify{25} fault types, each validated to produce meaningful user-perceivable impacts.

\subsubsection{Comparative Analysis of Benchmark Characteristics}
\label{sec:rq4_comparative_analysis}

Building on our curated set of high-impact failures, we now compare our benchmark against existing datasets across three key dimensions of complexity: workload intensity, system interaction coverage, and causal complexity.
Fig.~\ref{fig:rq4_dataset} provides a summary of this comparison.

\paragraph{Workload Intensity and Data Volume}

A benchmark's diagnostic challenge is fundamentally tied to its workload intensity, which dictates the volume and velocity of telemetry data generated.
We measure this intensity using Queries Per Second (QPS), derived from the total traces and duration.
Our benchmark operates at an average of \textbf{\modify{16.47} QPS} (max is \modify{37.6}), with fluctuations inherent to its dynamic request generation; we did not increase the QPS further to avoid system instability that would impact data generation efficiency.
This process generated a massive dataset of \textbf{\modify{154.7} million log lines} and \textbf{\modify{11.2} million traces} over \modify{188.1} hours.
This stands in contrast to existing benchmarks, whose intensity is often at a toy level.
For instance, Nezha-TT~\cite{nezhadataset}, Eadro-TT~\cite{eadrodataset},  RE2-TT and RE3-TT~\cite{pham2025rcaeval}, operate at a mere \textbf{\modify{0.14} QPS}, \textbf{\modify{0.25} QPS}, \textbf{\modify{4.35} QPS}, and \textbf{\modify{1.34} QPS}, respectively.
This low intensity fails to simulate the high-concurrency conditions where failures often emerge from complex interactions under load.
The large data volume in our benchmark is a direct consequence of this higher workload intensity, making it essential for assessing the scalability and noise-resistance of data-driven RCA models.

\paragraph{Breadth and Depth of System Interaction}
Beyond data volume, a benchmark's value is determined by its ability to exercise the system's architectural complexity.
The ability to trigger complex, cascading failures depends on how thoroughly the workload explores the microservice architecture.
We assess this using two key indicators: the service count, representing the system's potential scale, and the service coverage rate (the proportion of services exercised by a single trace), representing the interaction's breadth.
While some benchmarks like RE2-TT, Nezha-TT, and Eadro-TT span relatively large systems but still exhibit shallow workloads and low service coverage rates.
In contrast, our dynamic workload generator achieves a service coverage rate of \textbf{\modify{84\%}} (\modify{42} out of \modify{50} services), far exceeding RE2-TT's \textbf{\modify{39\%}} (\modify{27}/\modify{69}), Nezha-TT's \textbf{\modify{61\%}} (\modify{28}/\modify{46}) and Eadro-TT's \textbf{\modify{48\%}} (\modify{13}/\modify{27}).
This high coverage ensures that injected faults are not isolated but can propagate widely through the system's dependency graph, creating complex diagnostic challenges that require a holistic system view.

\paragraph{Complexity of Causal Propagation}
Finally, the ultimate test of an RCA benchmark is its ability to generate failures with long and complex causal chains.
We use the \textbf{maximum service call depth} as a direct proxy for a benchmark's potential causal complexity.
A deeper call chain allows for more intricate fault propagation paths and increases the likelihood of challenging phenomena like \textit{symptom drift}, where the most severe symptoms manifest far downstream from the actual root cause.
Our dataset achieves a maximum depth of \textbf{\modify{7}}  (corresponding trace length is 20), creating long and complex causal chains.
This is substantially deeper than most other benchmarks, which typically have depths between \modify{2} and \modify{4} (e.g., Nezha-OB: \modify{2}, Eadro-TT: \modify{3}).
The shallow propagation paths in datasets like GAIA (Max Depth of \modify{2}) mean the root cause is often coincident with the most obvious symptom, making it trivially identifiable and thus unsuitable for evaluating advanced causal inference capabilities.
Furthermore, our curated set of \textbf{\modify{78} unique metric types} provides the rich, multi-faceted signals necessary for complex diagnosis, balance between the sparse instrumentation of benchmarks like Eadro (\modify{7} types) and the potentially overwhelming noise of GAIA (\modify{949} types, most of the which are not relevant for RCA).

\begin{findingbox}{Our benchmark is qualitatively superior by design, creating complex diagnostic scenarios by combining high-impact failures, intense workloads, deep system interactions, and rich telemetry.}
\end{findingbox}

\input{figures/rq4_dataset.tex}

\subsection{RQ5: Performance Evaluation and Failure Mode Analysis}

Given the differences in characteristics observed between our benchmark and existing datasets, we first present the overall performance of \modify{eleven} SOTA RCA models on our new benchmark. 
From Fig.~\ref{fig:rq4_generation_process}, we see that the distribution of fault types in our benchmark is not balanced, simply inspecting the overall performance metrics can obscure important nuances in how different models handle various fault categories.
We then discuss the underlying reasons for their widespread failure, moving beyond simple metrics to a qualitative and quantitative analysis of their core limitations.

\subsubsection{Overall Performance}

We first evaluated all models on our entire benchmark and found overall low performance: the mean $Top@1$ accuracy across all models is only \textbf{0.21}, with the best-performing model, MicroRCA, reaching merely \textbf{0.37}. 
Our heuristic baseline, SimpleRCA, can also achieve only \textbf{0.28} $Top@1$ accuracy, further underscoring that the complex failure patterns in our dataset cannot be captured by simple rules and demand more sophisticated approaches.

\subsubsection{Performance Breakdown by Fault Type}
\input{figures/rq5_perf_by_fault_type.tex}

To understand this performance drop, we first break down the results by fault type, revealing that model effectiveness is highly dependent on the nature of the failure.
Fig.~\ref{fig:rq5_perf_by_fault_type} presents a detailed breakdown of $Top@1$, MRR, and $Avg@5$ for each model across various fault categories.
We choose these three metrics because $Top@1$ reflects the strictest accuracy requirement, MRR provides a balanced view of ranking quality, and $Avg@5$ offers a more stable and holistic view of ranking quality across a broader range of candidates.
From the figure, we observe several important patterns:
(1) \textbf{Performance varies significantly across different fault types, revealing the specific strengths and weaknesses of current RCA models.}
For instance, many models demonstrate high proficiency in diagnosing faults related to container lifecycle and resource stress, such as \texttt{ContainerKill}, \texttt{PodKill}, and \texttt{JVMMemoryStress}.
In these scenarios, models like Baro, SimpleRCA, and MicroRCA consistently achieve high scores (e.g., Baro's MRR is 0.87 for \texttt{ContainerKill} and 0.95 for \texttt{JVMMemoryStress}), likely because these faults produce clear, unambiguous signals (e.g., component crashes, resource usage spikes) that are easily captured by telemetry data.
(2) \textbf{Network-related and semantically complex faults pose a significant challenge to almost all models.}
Performance drops sharply for most network faults (\texttt{NetworkCorrupt}, \texttt{NetworkLoss}, etc.) and logic-related faults (\texttt{JVMReturn}).
For example, in \texttt{NetworkCorrupt} scenarios, even the best-performing model, Shapleyiq, only achieves an MRR of 0.40.
This suggests that current models struggle to diagnose problems that are subtle, manifest intermittently, or require a deep understanding of system interactions rather than just simple signal anomalies.
(3) \textbf{Certain fault types lead to near-total failure across all models.}
For faults like \texttt{DNSRandom} and \texttt{TimeSkew}, almost every model fails completely, with performance metrics at or near zero.
This highlights the existence of ``blind spots'' in current observability and modeling approaches, where the generated telemetry data may not contain sufficient information for diagnosis.
Interestingly, Shapleyiq shows exceptional performance on delay-based faults (\texttt{HTTPRequestDelay} and \texttt{HTTPResponseDelay}), achieving near-perfect scores (MRR > 0.97), indicating its specialized capability in handling latency-induced issues.

\subsubsection{Performance Breakdown by Algorithm}
\input{figures/rq5_perf_by_algo.tex}

Next, we analyze the results from the perspective of algorithmic families, which reveals distinct behavioral patterns in how different approaches rank potential root causes.
Fig.~\ref{fig:rq5_failure_distribution_by_fault_type} illustrates the performance distribution of each algorithm in terms of fault type, using $Top@1$ (\textcolor{topatonecolor}{red}), $Top@3$ (\textcolor{mrrcolor}{yellow}), and $Top@5$ (\textcolor{avgfivecolor}{green}) accuracy.
This perspective reveals distinct behavioral characteristics of different algorithmic approaches.
(1) \textbf{Some models exhibit an ``all-or-nothing'' diagnostic behavior.}
For algorithms like Eadro, MicroHecl, and to some extent MicroDig and MicroRCA, the bars for $Top@1$, $Top@3$, and $Top@5$ are often of the same height.
This indicates that if the model does not identify the correct root cause as its top-ranked candidate, it is highly unlikely to find it in the subsequent four results.
This suggests a lack of a nuanced ranking mechanism; the models' features either point directly to the correct cause or miss it entirely.
(2) \textbf{Other models demonstrate more effective ranking capabilities.}
In contrast, models like SimpleRCA, Shapleyiq, and Baro frequently show a significant gap between their $Top@1$ and $Top@5$ scores.
For example, in \texttt{HTTPRequestAbort} cases, SimpleRCA's $Top@1$ is only 0.13, but its $Top@5$ accuracy jumps to 0.93.
This pattern suggests that while these models may not always place the true root cause at the very top, their ranking logic is effective enough to include it within a small set of candidates, which is valuable in practical diagnostic scenarios.
(3) \textbf{Model performance is highly dependent on the fault category.}
No single algorithm excels across all fault types.
For example, Baro and SimpleRCA perform exceptionally well on lifecycle faults like \texttt{ContainerKill} and \texttt{PodKill}, but struggle with network issues.
Conversely, Shapleyiq dominates latency-based faults but performs poorly on many others.
This lack of a universally superior model underscores that the effectiveness of an RCA algorithm is tightly coupled to its underlying assumptions and the specific characteristics of the failure it is designed to detect.

The performance of SimpleRCA, in particular, serves as a crucial barometer for the complexity of our benchmark.
On our new, more challenging benchmark, its performance profile validates our design intentions.
The algorithm excels on faults like \texttt{ContainerKill} (MRR 0.75) and \texttt{JVMMemoryStress} (MRR 0.76) precisely because its design (aggregating simple, threshold-based anomaly counts) is perfectly suited for failures that produce a large volume of obvious signals (e.g., crashes, resource spikes) localized to a single component.
However, its effectiveness plummets on network-related and other subtle faults, where the root cause does not necessarily generate the most ``noise''.
\textbf{This demonstrates that our benchmark successfully creates complex scenarios where the true cause is decoupled from the most obvious symptoms, a condition that invalidates simple heuristic-based approaches.}
Therefore, SimpleRCA's performance pattern confirms that our benchmark introduces the kind of diagnostic ambiguity that necessitates the more sophisticated causal reasoning SOTA models claim to provide.

\subsubsection{Failure Mode Distribution}
\input{figures/rq5_scalability_collapse.tex}
To understand the root causes of this widespread performance degradation, we manually analyzed the \modify{262} most challenging cases where at least \modify{10} of the \modify{12} models failed (including SimpleRCA).
Our inductive coding process revealed three primary failure modes.
Since a single case could cause different algorithms to fail for different reasons, we allowed each case to be assigned multiple failure mode labels; the proportions shown in Fig.~\ref{fig:rq5_failure_distribution} are calculated from the total sum of these labels.
These modes represent a hierarchy of challenges, from practical viability to algorithmic soundness.

The most immediate barrier to adoption is \textbf{Scalability Issues}, which render many SOTA models impractical for production environments. 
Without resource limit, CausalRCA and Nezha need 927.11s and 94.62s on average to process our 8-minute data window (present in Table~\ref{tab:rq5_results_table}), respectively.
In our controlled scalability tests, we constrained all models to a realistic resource limit of 4 CPU cores.
As shown in Fig.~\ref{fig:rq5_scalability_collapse}, the execution time of most models increases at least linearly with data volume except for SimpleRCA. 
The time cost of Eadro and Art is very high, even with a small data volume,  since it needs parallelism to prepare the data format required by their deep learning models.

Beyond raw performance, even scalable models are frequently defeated by \textbf{Observability Blind Spots}, where necessary diagnostic signals are either absent or uninterpretable. We identified three distinct types: \textit{Signal-Loss Blind Spots}, where models fail to interpret the cessation of telemetry (e.g., from a \texttt{PodKill} fault) as a signal; \textit{Infrastructure-Level Blind Spots}, where faults occur in unmonitored components like databases, which we deliberately included to test robustness; and \textit{Semantic-Gap Blind Spots}, where contradictions between telemetry sources (e.g., an error log entry alongside a "success" trace status) create a paradox. These blind spots often compound, creating a web of contradictory signals that defeats models lacking the ability to reason under data uncertainty and incompleteness.

Ultimately, even with perfect scalability and complete observability, models fail due to inherent \textbf{Modeling Bottlenecks}, where their core assumptions are invalidated by complex failure patterns. A prime example is frequency-based models like Nezha~\cite{yu2023nezha}, which assume faults increase event frequency. This assumption is violated by faults like \texttt{PodKill} that cause signals to cease, leading the model to misinterpret the absence of events as normal operation. This highlights the most profound limitation: many models' rigid, assumption-laden approaches cannot accommodate the diverse and often subtractive nature of real-world failures, revealing a critical gap in their core diagnostic logic.

%% file: figures/rq4_generation_process.tex
\begin{figure}[t]
  \centering
  \includegraphics[width=1\textwidth]{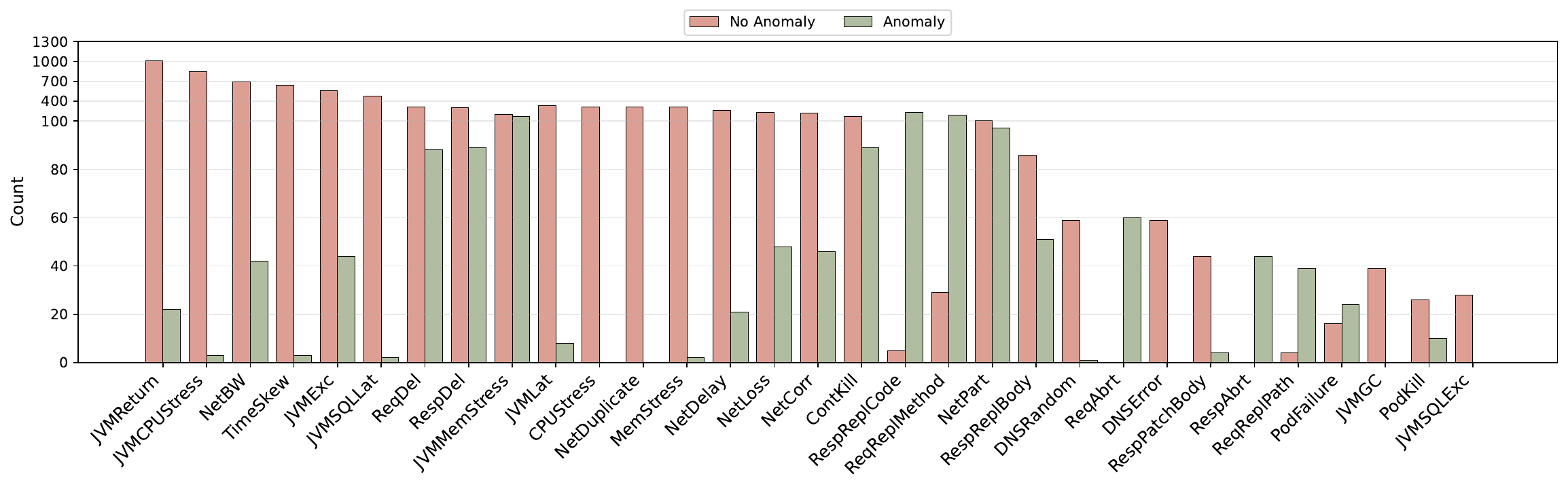}
  \caption{Three types of the datasets collected and detected in the dataset generation process in terms of the injected fault types. The labels are labeled by the detector introduced in Section~\ref{sec:impact_selection}, which detects the user-aware anomalies.}
  \label{fig:rq4_generation_process}
\end{figure}

%% file: figures/rq4_dataset.tex
\begin{figure}[t]
    \centering
    \begin{minipage}{0.48\textwidth}
        \centering
        \includegraphics[height=5cm]{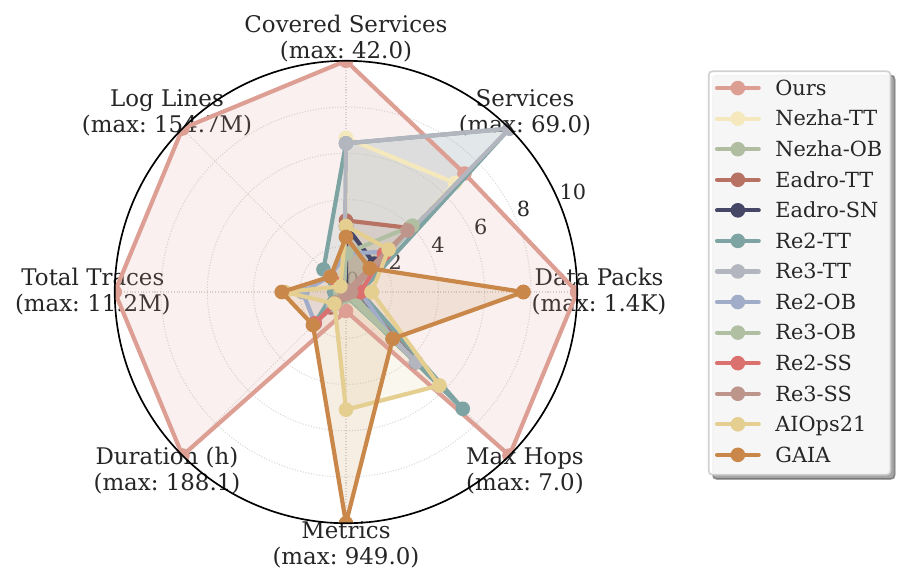}
        \caption{Dataset characteristics in terms of eight key attributes. Our newly constructed benchmark demonstrates a balanced and challenging profile across all dimensions, including request covered services, log numbers, request numbers, durations, and data-packs (distinct injected faults).}
        \label{fig:rq4_dataset}
    \end{minipage}
    \hfill
    \begin{minipage}{0.48\textwidth}
        \centering
        \footnotesize
        \captionsetup{type=table}
        \caption{Algorithm performance comparison on our newly constructed benchmark dataset. The best results are highlighted in \textbf{bold}.}
        \resizebox{\linewidth}{!}{%
        \begin{tabular}{@{}lccccccr@{}}
        \toprule
        Algorithm & $Top@1$ & $Top@3$ & $Top@5$ & Avg3 & Avg5 & MRR & Time (s) \\
        \midrule
        \multicolumn{8}{c}{\textit{Full Dataset}} \\
        \midrule
        Baro & 0.36 & 0.50 & 0.58 & 0.43 & 0.48 & 0.44 & 0.99 \\
        CausalRCA & 0.22 & 0.40 & 0.42 & 0.33 & 0.36 & 0.31 & 927.11 \\
        MicroDig & 0.35 & \textbf{0.61} & 0.67 & \textbf{0.49} & 0.56 & \textbf{0.48} & 19.03 \\
        MicroHECL & 0.34 & 0.34 & 0.34 & 0.34 & 0.34 & 0.34 & 34.82 \\
        MicroRank & 0.04 & 0.18 & 0.25 & 0.12 & 0.17 & 0.12 & 36.29 \\
        MicroRCA & \textbf{0.37} & 0.50 & 0.64 & 0.43 & 0.51 & 0.45 & 35.64 \\
        Nezha & 0.04 & 0.08 & 0.08 & 0.06 & 0.07 & 0.05 & 94.62 \\
        Shapleyiq & 0.22 & 0.36 & 0.52 & 0.29 & 0.37 & 0.31 & 42.74 \\
        SimpleRCA & 0.28 & 0.60 & \textbf{0.80} & 0.46 & \textbf{0.58} & 0.47 & \textbf{0.90} \\
        \midrule
        \multicolumn{8}{c}{\textit{Test Set}} \\
        \midrule
        DiagFusion & 0.13 & 0.17 & 0.22 & 0.15 & 0.17 & 0.16 & 3.03 \\
        Eadro & 0.16 & 0.16 & 0.16 & 0.16 & 0.16 & 0.16 & 32.75 \\
        Art & 0.04 & 0.08 & 0.10 & 0.06 & 0.07 & 0.06 & 8.10 \\
        \bottomrule
        \end{tabular}%
        }
        \label{tab:rq5_results_table}
    \end{minipage}
\end{figure}

%% file: figures/rq5_perf_by_fault_type.tex
\begin{figure}[t]
  \centering
  \includegraphics[width=1\textwidth]{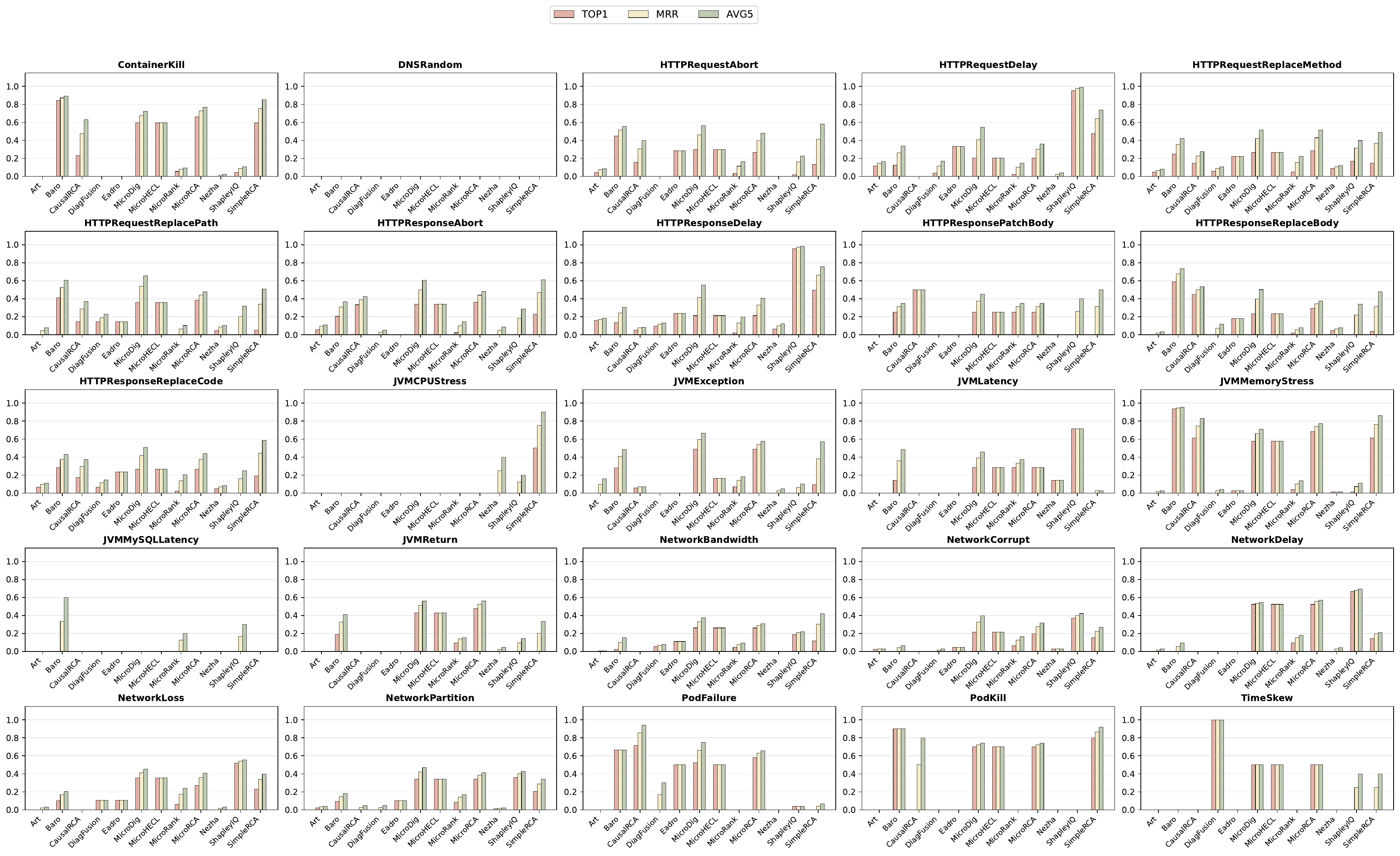}
  \caption{The performance distribution of each algorithm in terms of fault type, $Top@1$(\textcolor{topatonecolor}{red}), MRR(\textcolor{mrrcolor}{yellow}), $Avg@5$(\textcolor{avgfivecolor}{green}), respectively.  Most of  the algorithms perform poorly on the DNS failure, TimeSkew failure, CPU and Memory stress failure, as well as MySQL related failures.}
  \label{fig:rq5_perf_by_fault_type}
\end{figure}

%% file: figures/rq5_perf_by_algo.tex
\begin{figure}[t]
  \centering
  \includegraphics[width=1\textwidth]{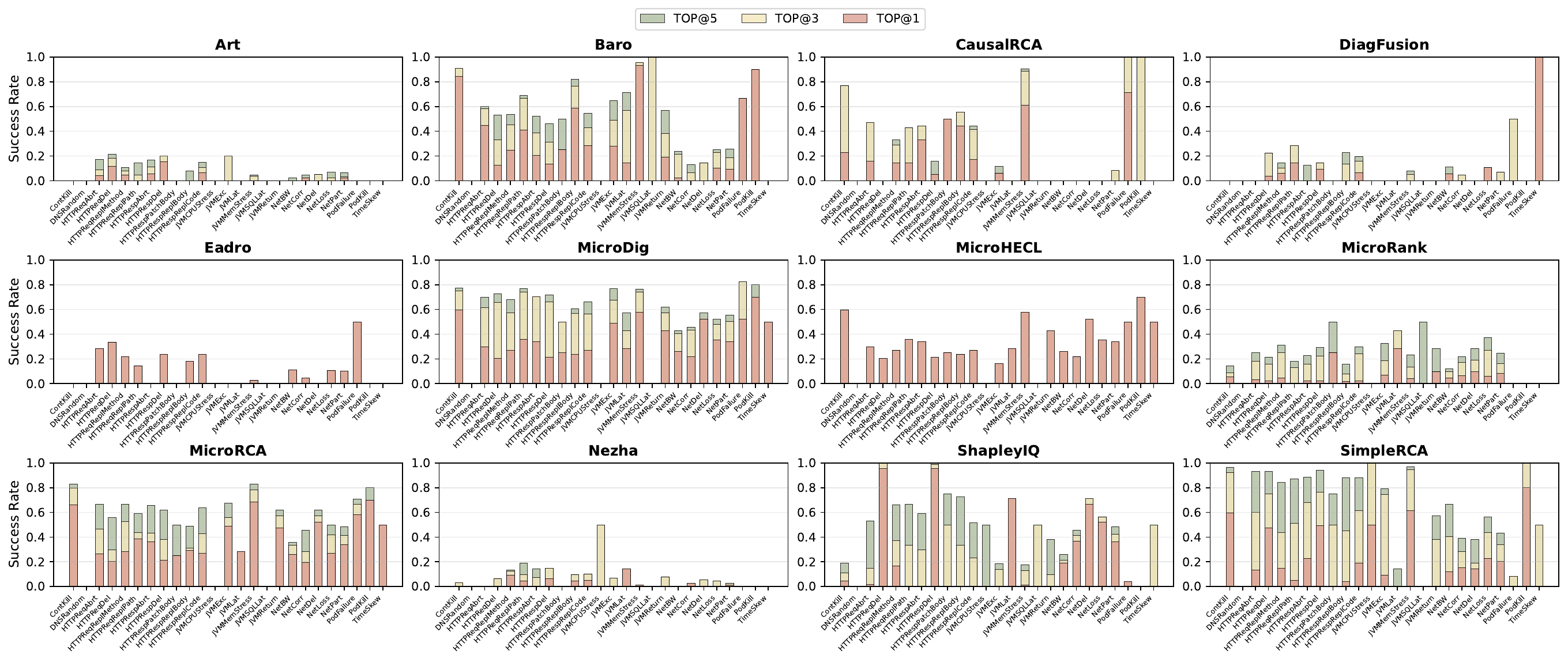}
  \caption{Failure rate distribution of each algorithm across different fault types. 
  The failure rate is measured at $Top@1$ (\textcolor{topatonecolor}{red}), $Top@3$ (\textcolor{mrrcolor}{yellow}), and $Top@5$ (\textcolor{avgfivecolor}{green}) accuracy levels.
  $Top@1$ shows the highest failure rate due to its strict accuracy requirement, while $Top@5$ demonstrates the lowest failure rate with relaxed criteria.
  When all bars appear green, it indicates that $Top@5$ and $Top@1$ achieve identical performance.}
  \label{fig:rq5_failure_distribution_by_fault_type}
\end{figure}

%% file: figures/rq5_scalability_collapse.tex

\begin{figure}[t]
    \centering
    \begin{minipage}{0.55\textwidth}
        \centering
        \includegraphics[width=1\textwidth]{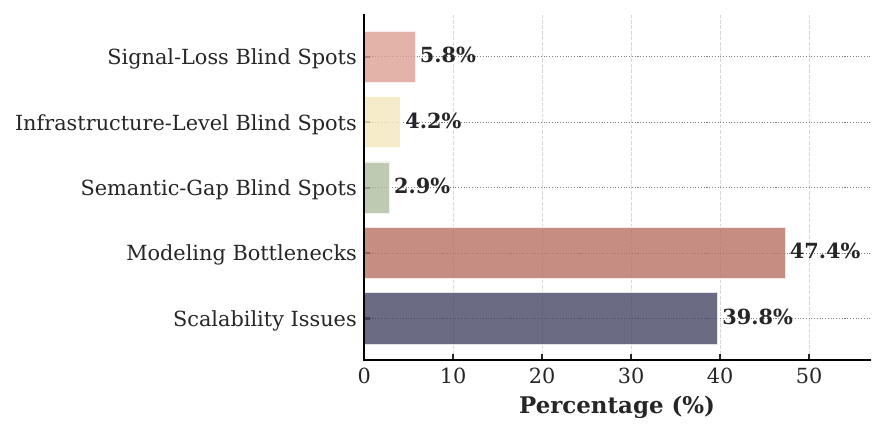}
        \caption{Distribution of failure causes across SOTA RCA algorithms in failure cases. 
        Failures are categorized into three main types: (1) observability blind spots where telemetry data is missing or incomplete, (2) modeling limitations in capturing system behaviors, and (3) scalability issues that prevent models from handling complex scenarios.}
        \label{fig:rq5_failure_distribution}
    \end{minipage}
    \hfill
    \begin{minipage}{0.43\textwidth}
        \centering
        \includegraphics[width=\textwidth]{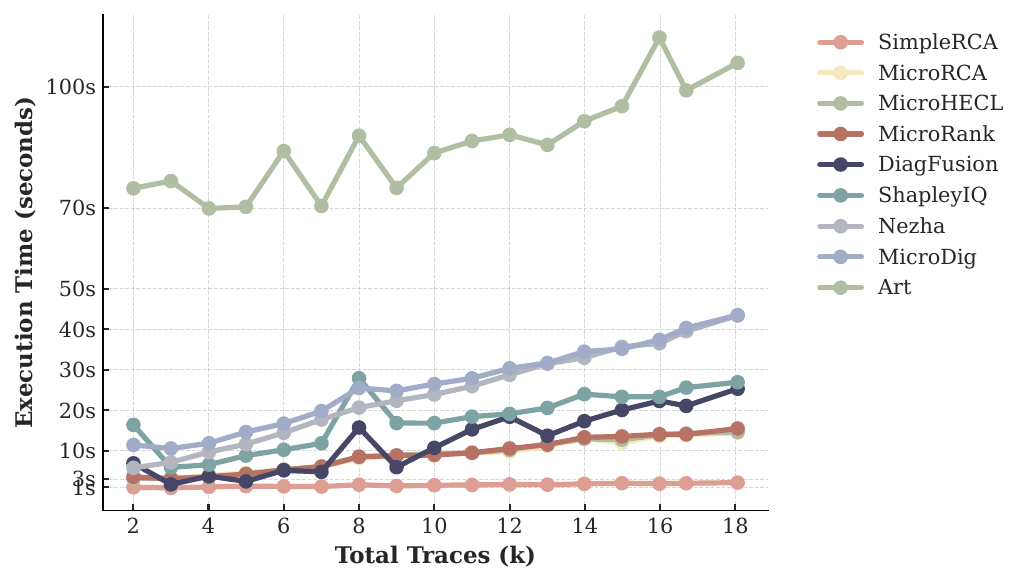}
        \caption{Scalability analysis of trace/log-based algorithms under 4-core CPU allocation. Most models exhibit at least linear time complexity, limiting real-time applicability.}
        \label{fig:rq5_scalability_collapse}
    \end{minipage}
\end{figure}

%% file: chapters/07discussion.tex
\section{Discussion}
\label{sec:discussion}

\subsection{Threats to Validity}
\label{sec:threats}

We analyze threats to our study's internal, external, and construct validity and our mitigation strategies.

\paragraph{Internal Validity}
This concerns the confidence in our causal conclusions.
A key threat is the configuration of baseline models; we mitigated this by using official implementations and prescribed setups.
Another threat is potential bugs in our experimental framework.
We addressed this through extensive testing and a feedback-guided data curation process, which provided continuous end-to-end validation of our toolchain.
Finally, to counter human error in our qualitative failure analysis, we established explicit classification criteria and had three authors independently review and cross-validate the results to ensure consensus.

\paragraph{External Validity}
This relates to the generalizability of our findings.
Our study relies on a single microservice system, Train-Ticket.
Although it is the largest open-source system of its kind, its architecture may not represent all production environments.
We mitigated this by injecting 31 diverse fault types and simulating dynamic workloads to capture challenges common to most microservice systems.
Our selection of eleven SOTA models may not cover the entire rapidly evolving field.
However, our analysis focuses on fundamental failure modes (e.g., Observability Blind Spots) likely shared by entire classes of models, making the findings broadly relevant.

\paragraph{Construct Validity}
This addresses whether our metrics accurately measure the intended concepts.
Our proxy metrics for benchmark complexity (e.g., QPS, service coverage) may not fully capture diagnostic difficulty.
We mitigated this by complementing quantitative metrics with a qualitative, in-depth failure analysis (RQ5).
Furthermore, standard evaluation metrics (e.g., Top@K, MRR) do not fully reflect the practical needs of an operator, such as the interpretability of results.
We acknowledge this limitation and scope our work to automated localization accuracy, a prerequisite for adoption, noting that full utility evaluation requires human studies.

\subsection{Implications for Future Research}
\label{sec:implications}

\subsubsection{Benchmark Design}
Our findings call for a significant evolution in benchmark design, moving beyond current limitations.
A primary challenge is the \textbf{oracle problem}, where reliance on user-facing SLIs creates a flawed ground truth.
This approach often mislabels subtle but genuine degradations as ``ineffective,'' which unfairly penalizes advanced models and underestimates their true diagnostic capabilities.
To address this, future benchmarks must broaden their \textbf{scope of failure} to include critical system-level issues, such as metastable states, rather than focusing only on user-perceivable incidents.
Furthermore, evaluation must evolve \textbf{beyond diagnostic accuracy}.
Standard metrics like \textit{Top@K} accuracy are insufficient because they ignore the practical costs of investigating ranked lists.
We advocate for new metrics that measure a model's \textbf{diagnostic coherence}, better reflecting its utility in real-world operations.

\subsubsection{Model Design}
Our failure analysis reveals critical directions for the next generation of RCA models.
The evidence shows that a ``one-size-fits-all'' approach is ineffective, highlighting a clear need for \textbf{fault-aware algorithm design} that tailors models to the distinct characteristics of different fault types.
In parallel, models must adopt a \textbf{holistic system modeling} perspective.
Instead of focusing on local, ``loudest noise'' anomalies, algorithms must be able to reason about cascading effects and systemic dependencies.
Finally, the \textbf{scalability-efficiency tradeoff} remains a major barrier to adoption.
Future work must prioritize lightweight architectures that balance high precision with the near-real-time diagnostic speeds required in production environments.

%% file: chapters/09conclusion.tex
\section{Conclusion}

In this paper, we examined the recent progress in data-driven RCA, questioning whether performance on existing benchmarks reflects true capability.
Our work suggests that the success of many SOTA models may be closely tied to the characteristics of current public benchmarks.
We supported this by analyzing these benchmarks, which revealed common limitations in workload intensity, interaction coverage, and causal complexity.
To address this gap, we developed an automated framework to generate a more comprehensive benchmark with dynamic workloads and a wider array of validated fault types.
Our re-evaluation of 11 SOTA models on this new benchmark showed that they achieve low \textit{Top@1} accuracies, averaging 0.21 with the best-performing model reaching merely 0.37.
This analysis helped us identify three common challenges for current methods: \textit{Scalability Issues}, \textit{Observability Blind Spots}, and \textit{Modeling Bottlenecks}.
Our findings suggest that many current RCA methods may face significant hurdles in complex production environments.
We propose that the research community could benefit from shifting focus from purely model-centric innovation to a co-design approach that also emphasizes the creation of more realistic and challenging benchmarks.
To facilitate this direction, we publicly release our benchmark, generation framework, and evaluation suite, providing a foundation for building the next generation of more robust diagnostic systems.

\section{Data Availability}

We promise that we will release all the code, baseline implementations, and datasets as open source.